\documentclass[preprint,11pt]{elsarticle}

\usepackage[margin=2.5cm]{geometry}
\usepackage{graphicx}
\usepackage{dcolumn}
\usepackage{xcolor}
\usepackage{bm}
\usepackage{amssymb}
\usepackage{amsmath}
\usepackage{algorithm}
\usepackage{algpseudocode}
\usepackage[english]{babel}
\usepackage{caption}
\usepackage{subcaption}

\begin{document}

\begin{frontmatter}


\title{\textcolor{black}{A probabilistic framework for irreversible kinetics}}

\author{Manas V. Upadhyay}
\affiliation{%
 organization={Laboratoire de M\'{e}canique des Solides (LMS), \'{E}cole Polytechnique, Institut Polytechnique de Paris, CNRS UMR 7649},
 addressline = {Route de Saclay},
 city={91128 Palaiseau},
 country={France}
}%


\begin{abstract}
A probabilistic framework for irreversible kinetics that unifies forward-in-time evolution and inverse inference, while quantifying epistemic uncertainty over admissible histories, is proposed. 
The forward-in-time evolution and inverse inference formulations differ only through observations and how they constrain admissible histories. 
All histories are ranked using a Gibbs-type probability measure $\exp(-\mathcal{J}/\Theta)$, where $\mathcal{J}$ encodes constitutive physics and observations. 
The parameter $\Theta$ controls epistemic uncertainty over the unknown histories, and the measure is interpreted as a Bayesian posterior over histories. 
Maximizing this posterior is equivalent to simultaneous minimization of $\mathcal{J}$ over the admissible history space. 
The framework is then specialized to generalized standard materials (GSM), for which $\mathcal{J}$ is constructed from incremental energy--dissipation functionals. 
This specialization reveals an important difference between the history resulting from simultaneous minimization over entire histories generally differs from the one generated by sequential GSM updates. 
Specifically, the causal GSM history incurs a higher history-level cost than the one obtained by simultaneous minimization, indicating that enforcing causality carries an additional cost at the history level.
This difference is illustrated through seven representative examples: viscoelasticity, viscoplasticity, rate-independent plasticity, gradient damage, conserved and non-conserved phase transformations, and coupled thermoelectric evolution. 
Finally, the finite-$\Theta$ formulation is demonstrated on an inverse problem with a non-convex free energy, where incomplete initial-state information is used to infer unobserved intermediate states and quantify uncertainty over admissible histories. 
As $\Theta$ decreases, the posterior concentrates towards the minimum-$\mathcal{J}$ histories.

\end{abstract}


\end{frontmatter}

\tableofcontents

\section{Introduction}\label{sec:intro}

\subsection{Motivation}\label{motivation}

\begin{figure}[h]
\centering
\includegraphics[width=\textwidth]{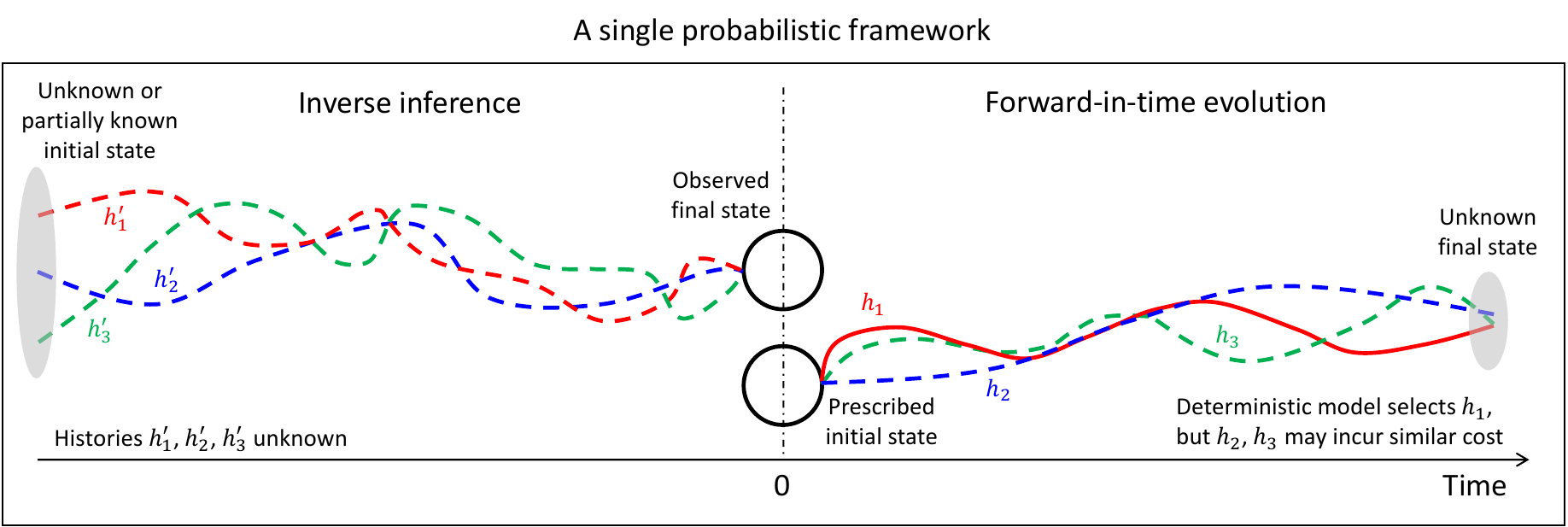}
\caption{\textcolor{black}{An illustration demonstrating the encapsulation of forward-in-time evolution and inverse inference of unknown admissible histories ($h_1$, $h_2$, $h_3$, $h_1'$, $h_2'$, $h_3'$) within the proposed single probabilistic framework.}}\label{fig:figure1}
\end{figure}

Irreversible kinetics is inherently path or history-dependent.
Even when the initial state of a body is known and the system is driven forward in time under prescribed loading, there may exist many irreversible paths that are consistent with the same available observations (Figure \ref{fig:figure1}). Such histories are referred to as \emph{admissible}. Admissibility does not imply that a history is a minimizer. A problem may possess a unique minimizing history while still admitting multiple other histories that satisfy the same physical and observational constraints.
In metastable systems, where several locally minimizing histories may coexist, this multiplicity can become more pronounced because each minimum may be surrounded by additional competing admissible histories.

The resulting epistemic uncertainty becomes even more pronounced when the problem is inverted.
Suppose instead that only the final state is given, possibly together with partial information about the initial state and without any knowledge of the preceding irreversible evolution (Figure \ref{fig:figure1}).
In such a situation, many admissible histories may be compatible with the available observations.
One may want to sample or characterize this set of histories and identify, under appropriate physics, the most probable histories.
This is fundamentally an inference problem over entire admissible histories and lies outside the scope of classical deterministic constitutive modeling.  

Furthermore, existing constitutive models are typically formulated as causal evolution laws, where each updated state is determined from the previously known state.
Such formulations are well suited to forward-in-time predictions, but they do not directly provide a framework for assigning probabilities to entire admissible histories or for inferring unknown past evolution from partial observations.

Therefore, there is a need to compare and rank the cost incurred by multiple admissible histories rather than considering a single deterministic constitutive path (see analogy in \ref{app:analogy}).
Such a ranking should allow alternative histories arising from incomplete information or unresolved microstructural variability to be compared according to their consistency with the prescribed physics and observations.

\subsection{Overview of the proposed probabilistic framework}

In this work, a probabilistic framework that compares and ranks admissible histories is proposed. It encompasses both (i) forward-in-time evolution of competing admissible histories and (ii) the inverse inference of unknown histories from sparse or partial observations (Figure \ref{fig:figure1}).
The two formulations use the same constitutive behavior and differ only through available observations and how they constrain admissible histories.

The development of this framework begins by associating each history $h$ in the continuous admissible history space $\mathcal{H}_{\rm ad}$, a cost through a constrained path functional $\mathcal{J}(h)$.
This functional encodes the constitutive physics and the constraints imposed by available observations that are not directly incorporated into the definition of $\mathcal{H}_{\rm ad}$.
A candidate set $\mathcal H\subseteq\mathcal H_{\rm ad}$ is then endowed with a discrete Gibbs-type probability measure proportional to $\exp(-\mathcal{J}\left(h\right)/\Theta)$, arising from Jaynes' maximum-entropy principle \cite{jaynesInformationTheoryStatistical1957, jaynesInformationTheoryStatistical1957a}, where $\Theta$ quantifies epistemic uncertainty over $\mathcal{H}$.
Apart from $\Theta$, the construction introduces no additional structure beyond that encoded in $\mathcal J$.

{At finite $\Theta$ and in the presence of observational constraints, the framework generates a posterior distribution that can be interpreted as a Bayesian posterior over admissible histories.
From this standpoint, it allows quantifying uncertainty in unobserved states, assigns relative probabilities to competing histories, and defines the set of maximum-a-posteriori (MAP) histories, through simultaneous maximization of the posterior probability, or equivalently, simultaneous minimization of $\mathcal{J}$ over $\mathcal{H}$.}
This discrete MAP problem is distinguished from simultaneous minimization over the continuous space $\mathcal H_{\rm ad}$.
Assuming that $\mathcal J$ is independent of $\Theta$, the exact discrete MAP set is also independent of $\Theta$.
In the limit $\Theta\to0$, the probability distribution concentrates on this set. Thus, $\Theta$ controls how broadly the probability is spread over admissible histories; at finite \(\Theta\), several histories can carry non-negligible probability, while as $\Theta \to 0$, the probability becomes concentrated on the minimum-$\mathcal{J}$ histories.

\color{black}
\subsection{Positioning and scope of this work}\label{sec:positioning}

The Gibbs-type probability measure over histories by itself is well established, and the closest related framework is the principle of maximum caliber (MaxCal).
In their review on MaxCal, Press\'e et al.~\cite{presse2013principles} have presented it as the extension of maximum-entropy inference from states to trajectories, where path entropy is maximized subject to prescribed dynamical constraints.
Dixit et al.~\cite{dixit2018maximum} have presented MaxCal as a general variational principle for inferring relative path weights from limited dynamical information.
Ge et al.~\cite{ge2012markov} have shown that particular choices of sequential pairwise constraints lead to Markov processes, illustrating the use of MaxCal to infer stochastic transition laws.
Path weights derived from MaxCal have also been used to sample trajectories of systems governed by deterministic Lagrangian dynamics \cite{gonzalezdiaz2020solving}.
The Gibbs-type measure is also structurally similar to Onsager-Machlup path measures for dissipative systems \cite{onsager1953} and to large-deviation path measures used in nonequilibrium thermodynamics \cite{grahamPathintegralMethodsNonequilibrium1978, freidlinRandomPerturbationsDynamical2012}.

Unlike MaxCal formulations, the proposed framework does not assume a pre-existing deterministic or stochastic equation of motion.
Instead, the proposed framework takes the constrained path functional $\mathcal J$, constructed directly from constitutive physics and available observations over the admissible history space, as the scalar cost defining the Gibbs-type measure.
Also, the framework does not require prescribing underlying stochastic or deterministic dynamics. 
Nevertheless, aleatory variability may be incorporated separately through stochastic inputs, observations, or latent variables when relevant.

In this regard, the framework itself is not tied to a particular class of constitutive models encoded in $\mathcal{J}$.
In the present work, it is demonstrated for incremental energy--dissipation cost built from a convex or non-convex free energy and a convex dissipation potential, corresponding to the class of generalized standard materials (GSM) \cite{halphen1975,ortiz1999,mielke2006,conti2008,mielke2008}, and more broadly, generalized gradient flow formulations \cite{ambrosioGradientFlowsMetric2008}.
Non-potential (antisymmetric, non-Onsager) flux-force couplings \cite{casimir1945,brechet2022,klika2024} and reversible-irreversible couplings of GENERIC type \cite{grmelaI1997,ottingerII1997} could be envisaged, but they are not treated in the present work.

Note that the use of the history-level functional $\mathcal{J}$ should not be interpreted as using a classical Hamiltonian variational principle, which relies on global-in-time variational structures and does not provide a universal description of nondissipative dynamics. 
In the GSM specialization considered in this work, irreversible evolution is represented through incremental energy--dissipation functionals of minimizing-movement type. 
This provides the constitutive structure required here; the more general problem of a unified spacetime variational formulation of reversible and irreversible dynamics could in principle be constructed within this framework, but such an extension is not developed in this work.

Within the GSM class of constitutive models, the same incremental energy--dissipation functionals lead to two generally different constructions for forward-in-time problems with a prescribed initial state: simultaneous optimization over the entire history and causal incremental minimization, the latter recovering the classical GSM evolution in the absence of observations.
The difference between these two constructions is examined in detail in Section~\ref{sec:theta0} through seven representative examples.
The proposed framework also allows the same GSM-type incremental energy--dissipation functionals to be applied to inverse inference, where final-state observations and partial information about earlier states are used to rank admissible histories.
In this sense, the proposed framework goes beyond a probabilistic reformulation of classical deterministic evolution by providing a single history-level formulation for both forward-in-time and inverse inference problems.

\subsection{Aims and manuscript structure}

The aims of this work are to formulate the proposed probabilistic framework, clarify the relationships among continuous minimizing, discrete MAP, and classical GSM histories, and demonstrate its applicability in forward-in-time evolution and inverse inference.

The general probabilistic framework and its $\Theta\to0$ limit are first established.
The framework is then specialized to GSM-type incremental energy--dissipation functionals, and the difference among the continuous minimizing history, the discrete MAP history, and causal GSM evolution is examined.
The seven representative forward-in-time examples are then presented to demonstrate that for each case the causal GSM history results in a higher $\mathcal{J}$ than the history selected by simultaneous optimization; these examples are linear viscoelasticity (Maxwell model), viscoplasticity (Norton model), rate-independent plasticity, gradient-regularized damage, non-conserved phase-field solidification (Allen-Cahn type), conserved phase-field solidification (Cahn-Hilliard type) and an Onsager-type coupled thermoelectric kinetics problem.
Finally, the finite-$\Theta$ inference capability with posterior statistics and uncertainty quantification are presented for a numerical problem involving a Swift-Hohenberg-type nonconvex energy and a prescribed final state with only partial information available on the initial state.

\section{\label{sec:framework}General probabilistic framework}

\subsection{Discrete Gibbs-type probability measure on admissible histories}\label{sec:gibbs}

Let $\mathcal H_{\rm ad}$ denote a continuous admissible history space, and let $\mathcal J:\mathcal {H}_{\rm ad}\rightarrow\mathbb R\cup\{+\infty\}$ be a given scalar constrained path functional encoding the constitutive physics and observations.
Let $\mathcal H\subseteq\mathcal H_{\rm ad}$ be an \textit{at most countable}\footnote{A discrete probability measure does not require the candidate set to be strictly finite as long as the candidate histories can be counted or enumerated and the partition function can be summed and remains positively finite, keeping the Gibbs probability measure well defined.} i.e., finite or countably infinite, candidate set. 
Here, admissibility implies belonging to $\mathcal H_{\rm ad}$, i.e., an admissible history satisfies the physical and observational requirements used to define this space, but it need not minimize $\mathcal J$ or belong to the minimizing set.

A discrete\footnote{A discrete Gibbs-type measure is adopted here for consistency with the intended numerical implementations. Defining a Gibbs-type measure directly on $\mathcal{H}_{\rm ad}$ would require, among other considerations, specifying an appropriate measurable structure and reference measure, establishing the existence and normalizability of the resulting measure, and replacing the discrete partition sum in \eqref{eqn:ZTheta} by an integral. Furthermore, the definition and variational characterization of MAP histories presented later would require separate treatment, particularly in an infinite-dimensional history space. Such a continuous-space extension is not considered here.} Gibbs-type probability measure is assigned to $\mathcal{H}$:
\begin{eqnarray}\label{eqn:gibbs}
    P_{\Theta}\left( h \right) = \frac{1}{Z_\Theta} \exp{\left( - \frac{\mathcal{J}(h)}{\Theta} \right)}, \quad h \in \mathcal{H},
\end{eqnarray}
and $P_\Theta(h) = 0$ for $h \notin \mathcal{H}$, with the discrete partition function
\begin{eqnarray}\label{eqn:ZTheta}
    Z_{\Theta} = \sum_{h \in \mathcal{H}} \exp{\left( - \frac{\mathcal{J}(h)}{\Theta} \right)},
\end{eqnarray}
where $\Theta > 0$ is the same temperature-like concentration parameter introduced in the introduction, and it is assumed that $0<Z_\Theta<\infty$.
$\Theta$ controls the spread of probability over $\mathcal H$ and therefore quantifies epistemic uncertainty over the candidate histories, and $P_{\Theta}$ represents the corresponding probability distribution over these histories.

Even though $\mathcal J$ is defined on $\mathcal H_{\rm ad}$, the Gibbs-type measure uses only its restriction to $\mathcal H$.
The framework then depends on the constitutive physics and observations only through $\mathcal H$ and the values of $\mathcal J$ on its elements.
Importantly, it does not depend on the particular functional decomposition used to construct $\mathcal J$.
In this sense, it is applicable to any constitutive model and observational constraints that can be appropriately encoded into $\mathcal J$ and the definition of $\mathcal H_{\rm ad}$.
It can distinguish between forward-in-time evolution and inverse inference through the available observations, as shall be shown later.

The Gibbs-type probability measure follows from Jaynes' maximum entropy principle \cite{jaynesInformationTheoryStatistical1957, jaynesInformationTheoryStatistical1957a}: among all probability distributions on $\mathcal H$ satisfying normalization and a prescribed value of the expectation of $\mathcal J$ i.e., $\mathbb E_\Theta(\mathcal J)$, it maximizes the Shannon entropy.
Assuming $\mathcal{J}$ is independent of $\Theta$, $\mathbb E_\Theta(\mathcal J)$ is finite, and differentiation under the sum is allowed, then $\mathbb{E}_{\Theta}(\mathcal{J})=\Theta^2\partial_\Theta\ln Z_\Theta$.

\subsection{Discrete MAP and continuous minimizing histories}\label{sec:MAP}

In Bayesian inference, a maximum a posteriori (MAP) estimate maximizes the posterior probability over the latent space.
In the present framework, the admissible history $h$ plays the role of the latent variable, and $P_\Theta(h)$ is interpreted as a posterior over $\mathcal H$.
The discrete MAP history set that maximizes $P_{\Theta}(h)$ is
\begin{equation}\label{eqn:MAPdef}
\mathcal M_{\rm MAP} := \underset{h\in\mathcal H}{\arg\max}\,P_\Theta(h).
\end{equation}
At this abstract level, no decomposition of $P_\Theta$ into prior and likelihood factors is assumed or required.
The probabilistic construction depends only on $\mathcal H$ and the values of the constrained path functional $\mathcal J$ over that set.

Since $Z_\Theta$ is independent of $h$ and $\exp(-\mathcal J(h)/\Theta)$ is strictly decreasing in $\mathcal J(h)$, the maximization of $P_\Theta(h)$ is equivalent to a minimization of $\mathcal{J}$ over $\mathcal{H}$ such that \eqref{eqn:MAPdef} becomes
\begin{equation}\label{eqn:MAPset}
\mathcal M_{\rm MAP} = \underset{h\in\mathcal H}{\arg\min}\, \mathcal J(h),
\end{equation}
and an element of this set is denoted by
\begin{equation}\label{eqn:globalh}
h_{\rm MAP}\in{\mathcal{M}}_{\rm MAP}.
\end{equation}

Meanwhile, assuming that the minimum of $\mathcal J$ over $\mathcal H_{\rm ad}$ is attained, the set of minimum-$\mathcal J$ continuous histories is defined as
\begin{equation}\label{eqn:Mcont}
\mathcal M_{\rm cont} := \underset{h\in\mathcal H_{\rm ad}}{\arg\min}\, \mathcal J(h),
\end{equation}
and an element of this set is denoted by
\begin{equation}\label{eqn:hcont}
h_{\rm cont}\in\mathcal M_{\rm cont}.
\end{equation}
Here, ``continuous'' refers to minimization over $\mathcal H_{\rm ad}$, and not to temporal continuity or discretization of the histories, both of which are permitted within the framework.

Since $\mathcal H\subseteq\mathcal H_{\rm ad}$,
the following condition is satisfied whenever both minima are attained,
\begin{equation}\label{eqn:JcontJMAP}
\mathcal J(h_{\rm cont}) \leq \mathcal J(h_{\rm MAP}).
\end{equation}
Equality holds if and only if $\mathcal{H}$ contains at least one continuous minimizer such that $\mathcal J(h_{\rm cont}) = \mathcal J(h_{\rm MAP})$, which is equivalent to $\mathcal M_{\rm cont}\cap\mathcal H\neq\varnothing$.
When this condition is satisfied,
$\mathcal M_{\rm MAP} = \mathcal M_{\rm cont}\cap\mathcal H$.
In particular, when the continuous minimizing history is unique and belongs to $\mathcal H$, $h_{\rm MAP}=h_{\rm cont}$.

All the stationarity conditions derived later characterize $h_{\rm cont}$ because they result from simultaneous minimization over the continuous space $\mathcal H_{\rm ad}$.
Generally, a discrete MAP history does not satisfy these stationarity conditions unless it is also a continuous minimizer.

Both $h_{\rm cont}$ and $h_{\rm MAP}$ are obtained by simultaneous minimizations\footnote{As discussed in section \ref{sec:positioning}, simultaneous optimization does not automatically imply that the formulation is symmetric upon time reversal. This becomes clearer with the GSM-type specialization in section \ref{sec:GSM}.} of $\mathcal{J}$ over entire histories, but crucially, they are performed over different sets.
The continuous history $h_{\rm cont}$ minimizes $\mathcal J$ over $\mathcal H_{\rm ad}$ and it is unique when $\mathcal M_{\rm cont}$ is a singleton.
Whereas the discrete MAP history $h_{\rm MAP}$ minimizes the restriction of $\mathcal J$ to $\mathcal H$, and it is unique when $\mathcal M_{\rm MAP}$ is a singleton.
These are distinct uniqueness statements because the corresponding minimizations are posed over different admissible history sets.

Simultaneous minimization of the entire history implies that the $h_{\rm cont}$ and $h_{\rm MAP}$ constructions, and equivalently $\mathcal{M}_{\rm cont}$ and $\mathcal{M}_{\rm MAP}$, are applicable to both forward-in-time evolution and inverse inference problems with the difference arising only from available observations encoded through $\mathcal{J}$ or in the definition of $\mathcal{H}_{\rm ad}$.

\subsection{Emergence of deterministic kinetics in the $\Theta \to 0$ limit}\label{sec:deterministic}

Assuming that the minimum of $\mathcal J$ over $\mathcal H$ is attained, let $\mathcal J_{\rm MAP}:= \min_{h\in\mathcal H}\mathcal J(h)$. The discrete MAP set is then $\mathcal{M}_{\rm MAP} = \{h \in \mathcal{H}| \mathcal{J}(h) = \mathcal J_{\rm MAP} \}$.
For any $h\in\mathcal{M}_{\rm MAP}$ and $h'\in\mathcal{H}\setminus\mathcal{M}_{\rm MAP}$,
\begin{equation}
\frac{P_\Theta(h')}{P_\Theta(h)}
=
\exp\left[
-\frac{\delta(h')}{\Theta}
\right],
\qquad
\delta(h')
:=
\mathcal{J}(h')-\mathcal{J}_{\rm MAP}>0.
\label{Xeqn7-9}
\end{equation}
Histories with $\mathcal J(h')=+\infty$ have zero probability for all $\Theta>0$.
As $\Theta\to0$, with $0<Z_\Theta<\infty$ for all $\Theta>0$, $P_\Theta(\mathcal M_{\rm MAP})\to1$ and $P_\Theta(\mathcal H\setminus\mathcal M_{\rm MAP})\to0$.

Assuming that $\mathcal J$ is independent of $\Theta$, $\mathcal M_{\rm MAP}$ is also independent of $\Theta$.
Then, the limit $\Theta\to0$ causes the probability to concentrate on $\mathcal M_{\rm MAP}$.
When $\mathcal M_{\rm MAP}$ contains a unique history, the probability concentrates on that history;
if multiple histories belong to $\mathcal M_{\rm MAP}$, then a unique deterministic history is not selected.

Whether this limiting MAP set also represents the continuous minimum depends on how \(\mathcal H\) samples \(\mathcal H_{\rm ad}\).
The selected deterministic history coincides with a continuous minimizing history when $\mathcal M_{\rm cont}\cap\mathcal H\neq\varnothing$.
Otherwise, the limit $\Theta\to0$ selects the history or histories with the lowest value of $\mathcal J$ represented within $\mathcal H$, rather than a minimizer over the full $\mathcal H_{\rm ad}$.

Note that the Bayesian interpretation, MAP set characterization, and $\Theta\to0$ concentration properties used above are standard consequences of a discrete Gibbs-type measure.
The novel contribution of this work lies in defining this measure directly from a constrained path functional $\mathcal J$ over admissible irreversible histories.

\section{Generalized standard materials (GSM)-type formulation of the probabilistic framework}\label{Xsec10-3}\label{sec:GSM}

The general probabilistic framework presented in section \ref{sec:framework} is specialized in this section using a GSM-type path action functional along with observational constraints. The resulting continuous minimizing history $h_{\rm cont}$ and the discrete MAP history $h_{\rm MAP}$ are distinguished from a causal incremental history $h_{\rm inc}$ that reduces to $h_{\rm GSM}$ in the absence of observations, corresponding to the sequence generated from classical GSM incremental updates.

For clarity, the developments below are done in a scalar setting; the same concepts directly extend to higher-order tensorial fields.

\subsection{Time-discretized constrained path functional}\label{Xsec11-3.1}\label{sec:incrementalaction}

A discrete set of times $\{t^0,\dots,t^N\}$ is chosen and a pair of fields $(z^n; q^n)$ is assigned to each time, where $z^n$ denotes internal variables describing irreversible changes and $q^n$ denotes prescribed quantities.

Let $\Delta t^n = t^{n+1} - t^n$.
The transition from $z^n$ to $z^{n+1}$ over $\Delta t^n$ is described by the following GSM incremental action functional representing an energetic and dissipative cost associated with the irreversible step under the prescribed field $q^{n+1}$:
\begin{equation}\label{eqn:Aincr}
\mathcal{A}^{n+1} \left(z^{n+1},z^n;q^{n+1}\right) := \Psi^{n+1} \left(z^{n+1},\nabla z^{n+1};q^{n+1}\right) + \Delta t^n \Phi^{n+1} \left(\frac{z^{n+1}-z^n}{\Delta t^n}\right)
\end{equation}
with $n = 0, \dots, N-1$.
Here,
\begin{equation}
\Psi^{n+1} \left(z^{n+1},\nabla z^{n+1};q^{n+1}\right) = \int_\Omega \psi^{n+1}\left(z^{n+1},\nabla z^{n+1};q^{n+1}\right) \,\mathrm{d}x
\label{Xeqn9-11}
\end{equation}
is the free energy functional of the domain $\Omega$, which may be non-convex in $z^{n+1}$ and may include gradient-regularization terms.
Whereas,
\begin{equation}
\Phi^{n+1}(r^{n+1}) = \int_\Omega \phi^{n+1}(r^{n+1})\,\mathrm{d}x, \label{Xeqn10-12}
\end{equation}
is the dissipation potential with $r^{n+1} = \left(z^{n+1}-z^n\right)/\Delta t^n$.
For rate-dependent models, it is assumed to satisfy coercivity conditions in the rate, whereas for rate-independent models it is positively homogeneous of degree one and may be non-smooth.

The incremental energy--dissipation action $\mathcal A^{n+1}$ is understood as an extended valued functional whose effective domain, including indicator functionals when appropriate, incorporates all constitutive and admissibility constraints, but no observational constraints.
In this regard, it has the standard variational structure associated with GSM models \cite{halphen1975,ortiz1999,mielke2006,conti2008,mielke2008}.
Each $\mathcal A^{n+1}$ is of minimizing-movement type and is oriented from the previous state $z^n$ towards the updated state $z^{n+1}$, reflecting the temporal direction of irreversible evolution.
Within this constitutive setting, thermodynamic consistency is imposed by requiring that the dissipation potential is proper, convex, lower semicontinuous, non-negative, and minimized at zero, with $\Phi^{n+1}(0)=0$.
Together with the associated constitutive flow rule, these assumptions ensure non-negative dissipation, but they are stronger assumptions than just satisfying the Clausius--Duhem inequality.

Each $z^n$ belongs to an appropriate state space $\mathcal Z$, chosen so that the energetic and dissipative functionals are well defined.
Any constitutive or admissibility restrictions on the internal variables are understood to be incorporated into $\mathcal Z$ or in the effective domain of $\mathcal A^{n+1}$.

Local observational contributions at time $t^n$ are collected separately into
\begin{equation}
\mathcal{C}^{n} \left(z^{n};q^{n} \right) = \int_\Omega c^{n} \left( z^{n};q^{n} \right) \text{ d}x, \quad n = 0, 1, \dots, N
\label{Xeqn11-13}
\end{equation}
with $q^{n}$ entering only as a prescribed quantity.
An initial state may be imposed through $\mathcal{C}^0$, while partial or endpoint observations may be imposed through the corresponding $\mathcal{C}^n$.
Alternatively, an exactly known initial state may be incorporated directly into the definition of the admissible set and $\mathcal{C}^0$ is omitted.
Hard observations may be represented by indicator functionals or incorporated directly into the admissible set, whereas imprecise observations may be introduced through finite penalty terms.

Observations connecting multiple times may also be prescribed through $\mathcal{C}^{\rm mult}_\alpha(z^{m_\alpha},\dots,z^{p_\alpha})$ with $0 \leq m_\alpha<p_\alpha$ and $\alpha\in\mathcal I$ (a finite index set).

The full continuous admissible history set for this problem is defined as
\begin{equation}\label{eqn:H}
\begin{aligned}
\mathcal{H}_{\rm ad} :=& \{ h=(z^0,\dots,z^N)\in\mathcal Z^{N+1} \big| (z^{n+1},z^n)\in\operatorname{dom}\mathcal A^{n+1}\ \forall n=0,\dots,N-1, \\
& \ \text{and all hard observational constraints imposed directly}
 \ \text{through the admissible set are satisfied} \},
\end{aligned}
\end{equation}
and $\mathcal H\subseteq\mathcal H_{\rm ad}$.

For a history $h\in\mathcal H_{\rm ad}$, the path action is defined as:
\begin{align}\label{eqn:action}
\mathcal{A} (h) := \sum_{n=0}^{N-1} \mathcal{A}^{n+1} \left(z^{n+1},z^n;q^{n+1} \right).
\end{align}

The corresponding observational constraints are
\begin{equation}\label{eqn:contraintfunc}
\mathcal{C}(h) := \sum_{n=0}^{N} \mathcal{C}^{n}(z^{n};q^{n}) + \sum_{\alpha \in \mathcal{I}} \mathcal{C}^{\rm mult}_\alpha (z^{m_\alpha},\dots,z^{p_\alpha}), \qquad 0 \leq m_\alpha < p_\alpha.
\end{equation}
$\mathcal{C}(h)$ respects the following conditions: When the initial state is prescribed directly through $\mathcal H_{\rm ad}$, and consequently through $\mathcal H$, the contribution $\mathcal C^0$ is omitted.
When all observational constraints are local in time, $\mathcal C^{\rm mult}_\alpha$ is absent.
Those hard observational constraints that are not imposed directly through $\mathcal{H}_{\rm ad}$ can be encoded by indicator function contributions in $\mathcal{C}(h)$, in which case violating histories have zero Gibbs-type weight.

The resulting GSM-type time-discretized constrained path functional is defined as
\begin{equation}\label{eqn:J}
\mathcal{J}(h) := \mathcal{A}(h) + \mathcal{C}(h).
\end{equation}
Note here that since the quantities $q^n$ are prescribed, their dependence is suppressed in the compact notation $\mathcal A(h)$, $\mathcal C(h)$, and $\mathcal J(h)$.

Defining the incremental constrained action functional
\begin{equation}\label{eqn:Jincr}
\mathcal J^{n+1}(z^{n+1},z^n;q^{n+1}) := \mathcal A^{n+1}(z^{n+1},z^n;q^{n+1}) + \mathcal C^{n+1}(z^{n+1};q^{n+1}),
\end{equation} the constrained path functional \eqref{eqn:J} can be expressed as
\begin{equation}\label{eqn:Jgeneral}
\mathcal J(h) = \mathcal C^0(z^0;q^0) + \sum_{n=0}^{N-1} \mathcal J^{n+1} (z^{n+1},z^n;q^{n+1}) + \sum_{\alpha \in \mathcal I} \mathcal C^{\rm mult}_\alpha (z^{m_\alpha},\dots,z^{p_\alpha}).
\end{equation}

Equations \eqref{eqn:Aincr}--\eqref{eqn:Jgeneral} provide the GSM-type construction of the abstract $\mathcal{H}_{\rm ad}$, $\mathcal{H}$ and $\mathcal{J}$ introduced in section \ref{sec:framework}.
This structure makes $\mathcal{A}$, and hence $\mathcal{J}$, fundamentally different from a classical Hamiltonian action.
Each incremental contribution is dissipative, first-order in time, and directed from the previous state towards the evaluated one. 

\subsection{Gibbs-type measure and simultaneous history optimization}\label{sec:gsm-gibbs}

For $h\in\mathcal H$, the Gibbs-type measure \eqref{eqn:gibbs} admits the following product representation\footnote{\textcolor{black}{This GSM-type construction also permits a prior--likelihood interpretation of \eqref{eqn:gibbs}. 
Let $y_{\rm obs}$ denote the available observations, and $h$ denote the latent history.
Using~\eqref{eqn:J}, $\exp\left(-\mathcal J/{\Theta}\right) = \exp\left(-\mathcal A/{\Theta}\right) \exp\left(-\mathcal C/{\Theta}\right)$.
One may identify $P_{\Theta}^{\rm prior}(h) \propto \exp\left[-\mathcal A(h)/{\Theta}\right]$ as a constitutive physics prior over admissible histories and
$P_{\Theta}(y_{\rm obs}\mid h) \propto \exp\left[-\mathcal C(h)/{\Theta}\right]$ as an observational likelihood. 
The posterior then satisfies $P_{\Theta}(h | y_{\rm obs}) \propto P_{\Theta}(y_{\rm obs}| h) P_{\Theta}^{\rm prior}(h) \propto \exp\left[-\mathcal J(h)/{\Theta}\right]$.
This prior--likelihood separation is one possible interpretation of the chosen additive decomposition of $\mathcal J$ in \eqref{eqn:J} and is not required for the general Gibbs-type construction of  section~\ref{sec:framework}. Importantly, this interpretation need not be unique. 
For example, hard observational constraints incorporated directly into the admissible set may equivalently be represented by indicator likelihood factors that vanish for histories violating those observations.}} for the GSM-type constrained path functional:
\begin{eqnarray}\label{eqn:multprob}
    P_\Theta(h) = \frac{1}{Z_\Theta} \exp\left(-\frac{\mathcal{C}^{0}(z^0;q^0)}{\Theta}\right) \prod_{n=0}^{N-1} \exp\left[-\frac{\mathcal{J}^{n+1}(z^{n+1},z^n;q^{n+1})}{\Theta}\right] \times \nonumber\\
    \prod_{\alpha \in \mathcal{I}} \exp\left[ -\frac{\mathcal{C}^{\rm mult}_\alpha (z^{m_\alpha},\dots,z^{p_\alpha})}{\Theta}\right].
\end{eqnarray}
If the initial state is prescribed directly through $\mathcal H_{\rm ad}$, and consequently through its candidate subset $\mathcal H$, then the factor involving $\mathcal C^0$ is omitted.
When observational constraints are only local in time, the final product term is omitted.

Note that even in the absence of the product involving $\mathcal C^{\rm mult}_\alpha$, the product on the right-hand side of \eqref{eqn:multprob} factorizes only the unnormalized history weight.
It does not imply statistical independence between increments or a product of normalized one-step conditional probabilities.
Adjacent increments share the intermediate states $z^n$, and normalization is performed only at the level of the entire history, with $Z_\Theta$ obtained by summing over all $h\in\mathcal H$.

Substituting \eqref{eqn:Jgeneral} for $\mathcal{J}(h)$ into the simultaneous minimization problem (\ref{eqn:globalh}) gives the MAP histories:
\begin{equation}\label{eqn:hMAP}
\begin{aligned}
h_{\rm MAP}\in \mathcal{M}_{\rm MAP} =  \underset{(z^0,\ldots,z^N)\in\mathcal H}{\arg\min} \Bigg[&\mathcal C^0(z^0;q^0) + \sum_{n=0}^{N-1} \mathcal J^{n+1}(z^{n+1},z^n;q^{n+1})
+ \sum_{\alpha\in\mathcal I} \mathcal C_\alpha^{\rm mult} (z^{m_\alpha},\dots,z^{p_\alpha})\Bigg],
\end{aligned}
\end{equation}
with the same conditions for uniqueness as those for \eqref{eqn:globalh} and omission of $\mathcal{C}^0$ and $\mathcal{C}^{\rm mult}_\alpha$ as for \eqref{eqn:multprob}.

The corresponding continuous minimizing history is obtained by posing the same simultaneous minimization problem over $\mathcal H_{\rm ad}$:
\begin{equation}\label{eqn:hcontGSM}
\begin{aligned}
h_{\rm cont}\in \mathcal{M}_{\rm cont} =  \underset{(z^0,\ldots,z^N)\in\mathcal H_{\rm ad}}{\arg\min} \Bigg[ &\mathcal C^0(z^0;q^0) +\sum_{n=0}^{N-1} \mathcal J^{n+1}(z^{n+1},z^n;q^{n+1})
+ \sum_{\alpha\in\mathcal I} \mathcal C_\alpha^{\rm mult} (z^{m_\alpha},\dots,z^{p_\alpha}) \Bigg].
\end{aligned}
\end{equation}
The minimization problems \eqref{eqn:hMAP} and \eqref{eqn:hcontGSM} have the same functional structure but are posed over the distinct history sets $\mathcal H$ and $\mathcal H_{\rm ad}$, respectively.

Assuming that the relevant variations are admissible, or that hard constraints have been incorporated into $\mathcal J$ through indicator functionals, and under the assumptions required for the corresponding subdifferential sum, any $h_{\rm cont}$ satisfies the necessary condition on interior states $z^{n+1}$:
\begin{equation}\label{eqn:stationaryadj}
\begin{aligned}
0\in{}& \partial_{z^{n+1}} \mathcal J^{n+1}(z^{n+1},z^n;q^{n+1}) + \partial_{z^{n+1}} \mathcal J^{n+2}(z^{n+2},z^{n+1};q^{n+2})
+ \sum_{\substack{\alpha\in\mathcal I\\ m_\alpha\leq n+1\leq p_\alpha}} \partial_{z^{n+1}} \mathcal C_\alpha^{\rm mult} (z^{m_\alpha},\dots,z^{p_\alpha}),
\end{aligned}
\end{equation}
for $n=0,\dots,N-2$.
Here, $\partial$ denotes an appropriate generalized subdifferential for the adopted variational setting; in the smooth case, \eqref{eqn:stationaryadj} reduces to the corresponding classical stationarity condition.
This stationarity condition characterizes the continuous minimization problem over $\mathcal H_{\rm ad}$ and is therefore a necessary condition for $h_{\rm cont}$ under the stated assumptions.
In general, it is not a necessary condition for the discrete MAP history $h_{\rm MAP}$, unless that history is also a continuous minimizer.

For the final state $z^N$, the necessary condition is
\begin{equation}
0\in \partial_{z^N} \mathcal J^N(z^N,z^{N-1};q^N) + \sum_{\substack{\alpha\in\mathcal I\\
m_\alpha\leq p_\alpha = N }} \partial_{z^N} \mathcal C_\alpha^{\rm mult} (z^{m_\alpha},\dots,z^{p_\alpha}).
\label{Xeqn20-24}
\end{equation}

If $z^0$ is unknown, its necessary condition is
\begin{equation}\label{eqn:stationaryadjinit}
0\in \partial_{z^0}\mathcal C^0(z^0;q^0) + \partial_{z^0} \mathcal J^1(z^1,z^0;q^1) + \sum_{\substack{\alpha\in\mathcal I\\m_\alpha=0 < p_\alpha}} \partial_{z^0} \mathcal C_\alpha^{\rm mult} (z^{m_\alpha},\dots,z^{p_\alpha}).
\end{equation}

Note that the stationarity conditions \eqref{eqn:stationaryadj}--\eqref{eqn:stationaryadjinit} characterize $h_{\rm cont}$; they characterize $h_{\rm MAP}$ only when that history is also a continuous minimizer.

Similar to the discussion in section \ref{sec:gibbs}, the simultaneous minimization of complete histories in \eqref{eqn:hMAP} and \eqref{eqn:hcontGSM} does not make the formulation symmetric upon time reversal.
Each contribution to $\mathcal A$ remains causal.
In the continuous minimization problem \eqref{eqn:hcontGSM}, simultaneous optimization yields stationarity conditions \eqref{eqn:stationaryadj} for each interior state $z^{n+1}$ that involve contributions from both the preceding increment $z^n$ through $\mathcal J^{n+1}(z^{n+1},z^n;q^{n+1})$, subsequent increment $z^{n+2}$ through $\mathcal J^{n+2}(z^{n+2},z^{n+1};q^{n+2})$, and every $\mathcal{C}^{\rm mult}_\alpha$ whose argument contains $z^{n+1}$, even though the underlying incremental constitutive structure remains irreversible and causal.
Therefore, simultaneous optimization over the unknown states of the complete history does not make the formulation time-symmetric or equivalent to a reversible Hamiltonian variational principle.
At the same time, however, in the forward-in-time evolution case, simultaneous minimization of $\mathcal J$ over the entire history is not equivalent to the sequential incremental minimization defining the causal GSM evolution; this is developed in the next section.

\subsection{Causally generated incremental histories for forward-in-time evolution: Connection with GSM updates}\label{Xsec13-3.3}\label{sec:forward}

A causal incremental history $h_{\rm inc}$ is constructed here by sequentially maximizing the unnormalized incremental Gibbs-type factors associated with the same GSM-type energy--dissipation functional $\mathcal A^{n+1}$ \eqref{eqn:Aincr} together with local observational contributions.
When observations are absent, $h_{\rm inc}$ is equal to the GSM history $h_{\rm GSM}$.
For both constructions, the initial state is prescribed, and no constraints coupling multiple times are included; consequently, these constructions are only applicable to forward-in-time evolution.

With $z^0$ prescribed and observational constraints being local in time, the history-level probability \eqref{eqn:multprob} can be written as
\begin{equation}\label{eqn:factorizedHistory}
P_\Theta\left(h\mid z^0\right) = \frac{1}{Z_\Theta(z^0)} \prod_{n=0}^{N-1} \exp\left[-\frac{ \mathcal J^{n+1}\left(z^{n+1},z^n;q^{n+1}\right) }{\Theta}\right], \qquad h\in\mathcal H(z^0),
\end{equation}
where $\mathcal{H}(z^0) \subseteq \mathcal{H}_{\rm ad}(z^0)$ with $\mathcal{H}_{\rm ad}(z^0)$ being $\mathcal{H}_{\rm ad}$ with the prescribed $z^0$ incorporated directly into the definition \eqref{eqn:H}, and the normalization factor $Z_\Theta(z^0)$ is
\begin{equation}
Z_\Theta(z^0) = \sum_{h=(z^0,\dots,z^N)\in\mathcal H(z^0)} \prod_{n=0}^{N-1} \exp\left[-\frac{\mathcal J^{n+1}(z^{n+1},z^n;q^{n+1})}{\Theta}\right].
\label{Xeqn23-27}
\end{equation}

At step $n$, the previously determined state $z_{\mathrm{inc}}^n$ is held fixed, and the next state $z^{n+1}_{\rm inc}$ is selected by maximizing only the corresponding incremental factor,
\begin{equation}
w_\Theta^{n+1} \left(z^{n+1};z_{\mathrm{inc}}^n\right) := \exp\left[-\frac{\mathcal J^{n+1} \left(z^{n+1},z_{\mathrm{inc}}^n;q^{n+1}\right) }{\Theta} \right].
\label{Xeqn24-28}
\end{equation}
This factor is not a separately normalized probability distribution.
No incremental partition function is required here because with a fixed $z_{\mathrm{inc}}^n$, such a partition function would not affect the maximizing state.
Importantly, the incremental optimization is not restricted to the discrete candidate set $\mathcal H(z^0)$.
Even though the factors in equation~\eqref{eqn:factorizedHistory} are evaluated on the discrete candidate histories, the same exponential form is used below to optimize $z^{n+1}$ over the continuous set of admissible next states.
Consequently, the resulting history $h_{\rm inc}$ need not belong to $\mathcal H(z^0)$ and should not be interpreted as a discrete MAP history.

For the corresponding continuous incremental variational problem, let $\mathcal Z_{\rm ad}^{n+1}(z_{\rm inc}^n;q^{n+1})$ denote the set of admissible next states satisfying the constitutive physics, and any observational constraints local to
time $t^{n+1}$.
Then,
\begin{equation}\label{eqn:zinc}
\begin{aligned}
z_{\mathrm{inc}}^{n+1} \in& \underset{z^{n+1}\in
\mathcal Z_{\rm ad}^{n+1}(z_{\rm inc}^n;q^{n+1})}{\arg\max} w_\Theta^{n+1} \left(z^{n+1};z_{\mathrm{inc}}^n\right) \\
&= \underset{z^{n+1}\in \mathcal Z_{\rm ad}^{n+1}(z_{\rm inc}^n;q^{n+1})}{\arg\min} \mathcal J^{n+1} \left(z^{n+1},z_{\mathrm{inc}}^n;q^{n+1}\right).
\end{aligned}
\end{equation}
Assuming that $\mathcal J^{n+1}$ is independent of $\Theta$, these argmax and argmin sets coincide for every $\Theta>0$.
Hence, they are independent of $\Theta$.

Under the assumptions required for the corresponding subdifferential optimality condition, a necessary condition for the incremental minimization problem is
\begin{equation}\label{eqn:stationaryinc}
0 \in \partial_{z^{n+1}} \mathcal J^{n+1} \left(
z_{\mathrm{inc}}^{n+1},z_{\mathrm{inc}}^{n};q^{n+1}\right).
\end{equation}

If the incremental admissible set is convex and the incremental constrained functional $\mathcal J^{n+1}$ is strictly convex over this set, then $z_{\rm inc}^{n+1}$, if it exists, is the unique incremental minimizer.
For smooth energy, dissipation, and observational contributions, and in the absence of active hard constraints, the subdifferential inclusion \eqref{eqn:stationaryinc} yields the following bulk stationarity equation in the domain $\Omega$, together with the natural or prescribed boundary conditions associated with the adopted admissible space:
\begin{equation}\label{eqn:zinc2}
\begin{aligned}
0 ={}& \left(\partial_z-\nabla\cdot\partial_{\nabla z} \right)\psi^{n+1} \left(z_{\mathrm{inc}}^{n+1},\nabla z_{\mathrm{inc}}^{n+1};q^{n+1} \right)
+\partial_r\phi^{n+1} \left(r_{\mathrm{inc}}^{n+1}\right)+\partial_z c^{n+1}\left( z_{\mathrm{inc}}^{n+1};q^{n+1}\right),\end{aligned}
\end{equation}
where
\begin{equation}
r_{\mathrm{inc}}^{n+1} := \frac{ z_{\mathrm{inc}}^{n+1} - z_{\mathrm{inc}}^n}{\Delta t^n}.
\label{Xeqn28-32}
\end{equation}

Repeating this incremental minimization for $n=0,\ldots,N-1$ generates the causal history
\begin{equation}
h_{\rm inc} = (z_{\rm inc}^0,\ldots,z_{\rm inc}^N) \in\mathcal H_{\rm ad}(z^0),
\label{Xeqn29-33}
\end{equation}
This is a sequential construction of incrementally optimized states and not a simultaneous minimization over $\mathcal H(z^0)$.
No separate incremental probability measure is introduced.

When the observational contributions vanish and no observational restrictions remain incorporated into the incremental admissible set or the effective domain of $\mathcal J^{n+1}$, equation~\eqref{eqn:zinc} reduces to the classical GSM incremental minimization problem \cite{halphen1975,ortiz1999,mielke2006,conti2008,mielke2008}. 
Furthermore, under the aforementioned convexity conditions, equation~\eqref{eqn:zinc2} reduces to the standard stationarity condition associated with the GSM incremental update making $z^{n}_{\rm inc} = z^n_{\rm GSM}$.
Finally, if the initial state is the same and the incremental minimizer is unique at every step, sequential application of these updates gives
\begin{equation}\label{eqn:hincGSM}
h_{\rm inc}=h_{\rm GSM}.
\end{equation}
Therefore, the deterministic GSM evolution is recovered directly through causal incremental minimization.
This is consistent with established variational formulations for rate problems and constitutive updates \cite{ortiz1999,mielke2006,conti2008}.

If some incremental minimizers are non-unique, the two constructions instead generate the same set of admissible causal histories, but a particular history is not uniquely selected without an additional selection rule.

\subsection{Relationships among continuous minimizing, discrete MAP, causal incremental, and GSM histories for the forward-in-time specialization}\label{Xsec14-3.4}

Let $z_{\rm inc}^0:=z^0$.
When $h_{\rm inc}\in\mathcal H(z^0)$, the history-level probability assigned to it by \eqref{eqn:factorizedHistory} is
\begin{equation}
P_\Theta\left(h_{\rm inc}\mid z^0\right) = \frac{1}{Z_\Theta(z^0)} \prod_{n=0}^{N-1} \exp\left[ -\frac{ \mathcal J^{n+1} \left(z_{\rm inc}^{n+1},z_{\rm inc}^n;q^{n+1}\right) }{\Theta} \right],
\label{Xeqn31-35}
\end{equation}
and does not imply that $h_{\rm inc}$ maximizes \eqref{eqn:factorizedHistory}.
In general, sequential maximization of the individual incremental factors does not maximize their product over complete histories.
Therefore, for any $h_{\rm MAP}$ obtained by maximizing \eqref{eqn:factorizedHistory} over the candidate set $\mathcal H(z^0)$,
\begin{equation}
P_\Theta(h_{\rm MAP}\mid z^0) \geq P_\Theta(h_{\rm inc}\mid z^0),
\label{Xeqn32-36}
\end{equation}
whenever $h_{\rm inc}\in\mathcal H(z^0)$.
Equivalently,
\begin{equation}
\mathcal J(h_{\rm MAP}) \leq \mathcal J(h_{\rm inc}).
\label{Xeqn33-37}
\end{equation}
Thus, even though the same GSM-type incremental functionals enter both constructions, $h_{\rm inc}$ and $h_{\rm MAP}$ are obtained through different optimization procedures and, in general, $h_{\rm inc}\neq h_{\rm MAP}$.

The relationship between $h_{\rm inc}$ and $h_{\rm cont}$ can be examined using the continuous history-level optimality conditions.
In the absence of observational constraints, for every interior state, the continuous minimizing history $h_{\rm cont}$ satisfies
\begin{equation}\label{eqn:stationaryadjforward}
\begin{aligned}
0\in{}& \partial_{z^{n+1}} \mathcal J^{n+1} \left(z^{n+1},z^n;q^{n+1}\right)
+ \partial_{z^{n+1}} \mathcal J^{n+2} \left(z^{n+2},z^{n+1};q^{n+2}\right), \qquad n=0,\ldots,N-2.
\end{aligned}
\end{equation}

Meanwhile, the incremental construction satisfies
\begin{equation}
0\in \partial_{z^{n+1}} \mathcal J^{n+1} \left(z_{\rm inc}^{n+1},z_{\rm inc}^{n};q^{n+1} \right), \qquad n = 0,\dots,N-1.
\label{Xeqn35-39}
\end{equation}
Consequently, the additional strong compatibility condition
\begin{equation}\label{eqn:compatibility}
0\in \partial_{z^{n+1}} \mathcal J^{n+2} \left( z_{\rm inc}^{n+2},z_{\rm inc}^{n+1};q^{n+2}\right), \qquad n=0,\ldots,N-2,
\end{equation}
is sufficient for $h_{\rm inc}$ to satisfy the interior stationarity conditions of $h_{\rm cont}$.
The final-state condition is already satisfied by the final incremental update, while no stationarity condition is imposed with respect to the prescribed initial state.

If, in addition to \eqref{eqn:compatibility}, $\mathcal H_{\rm ad}(z^0)$ is convex and the complete constrained path functional $\mathcal J$ is convex over this set, these conditions are sufficient for
\begin{equation}
h_{\rm inc}\in\mathcal M_{\rm cont}.
\label{Xeqn37-41}
\end{equation}
If $\mathcal J$ is strictly convex over $\mathcal H_{\rm ad}(z^0)$, then the continuous minimizer is unique and
\begin{equation}
h_{\rm inc}=h_{\rm cont}.
\label{Xeqn38-42}
\end{equation}

Therefore, without the required convexity, satisfaction of \eqref{eqn:compatibility} provides only a stationary history and does not establish that $h_{\rm inc}$ is a continuous minimizer over entire history.

The relationship with the discrete MAP history follows separately.
If $h_{\rm inc}\in \mathcal M_{\rm cont}\cap\mathcal H(z^0)$
then, by the conditions established in section~\ref{sec:MAP}, $h_{\rm inc}\in\mathcal M_{\rm MAP} = M_{\rm cont}\cap\mathcal H(z^0)$.
In particular, if the continuous minimizer is unique and belongs to $\mathcal H(z^0)$, then
\begin{equation}\label{eqn:hMAPcontinc}
h_{\rm cont}=h_{\rm MAP}=h_{\rm inc}.
\end{equation}
More generally, $h_{\rm inc}$ may be a discrete MAP history without being a continuous minimizer if it belongs to $\mathcal H(z^0)$ and minimizes $\mathcal J$ over that candidate set.
Thus, the continuous stationarity and the strong compatibility conditions are not necessary conditions for equality between $h_{\rm inc}$ and a discrete MAP history.

Finally, in the highly specialized case when observational contributions and restrictions on the admissible sets are absent, the causal incremental and classical GSM problems have the same incremental minimizers, and all the other conditions leading to equation~\eqref{eqn:hMAPcontinc} are satisfied, then
\begin{equation}\label{eqn:hMAPcontincGSM}
h_{\rm cont} = h_{\rm MAP} = h_{\rm inc} = h_{\rm GSM}.
\end{equation}

\begin{figure}
\centering
\includegraphics[width=\textwidth]{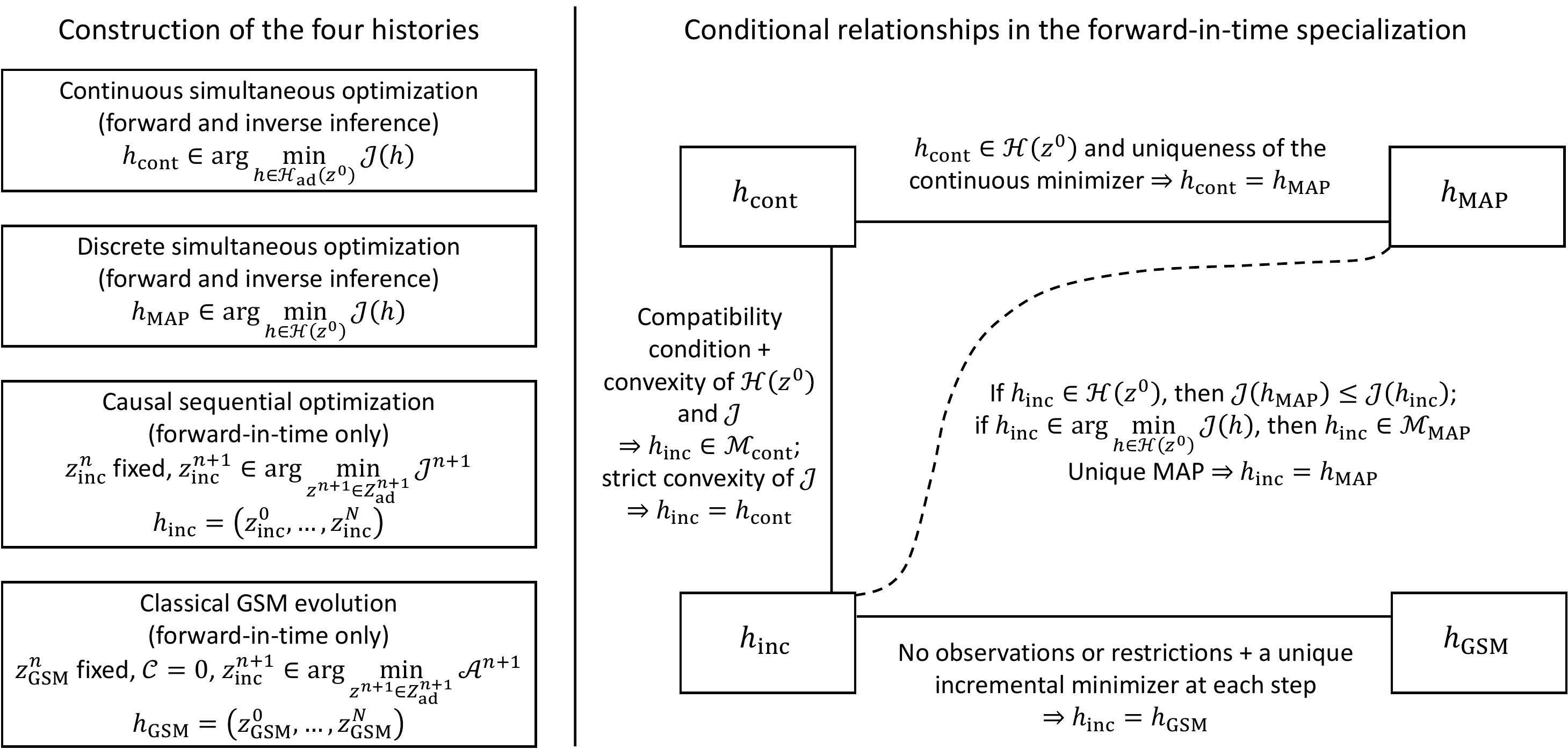}
\caption{Relationships among $h_{\rm cont}$, $h_{\rm MAP}$, $h_{\rm inc}$, and $h_{\rm GSM}$ under forward-in-time evolution. By construction, $h_{\rm cont}$ minimizes $\mathcal J$ over $\mathcal H_{\rm ad}$.}
\label{fig2}
\end{figure}

For the non-trivial evolutions considered in Section~\ref{Xsec16-4}, the strong compatibility condition \eqref{eqn:compatibility} is generally not satisfied. 
Consequently, the causal GSM history does not coincide with the simultaneously minimizing history. 
Since the latter minimizes $\mathcal J$ over the entire admissible history space, the causal construction generally incurs a higher history-level cost.

\subsection{Deterministic kinetics in the $\Theta \to 0$ limit}\label{Xsec15-3.5}\label{sec:gsm-deterministic}

The $\Theta \to 0$ limit established in section~\ref{sec:deterministic} applies directly to the discrete Gibbs-type measure associated with the GSM-type constrained path functional.
Assuming $\mathcal J$ is independent of $\Theta$, this limit causes the probability to concentrate on $\mathcal M_{\rm MAP}$.

Meanwhile, the causal incremental construction of section~\ref{sec:forward} does not require a $\Theta \to 0$ limit.
For every $\Theta>0$,
\begin{equation}
\arg\max w_\Theta^{n+1}
=
\arg\min \mathcal J^{n+1},
\label{Xeqn41-45}
\end{equation}
provided that $\mathcal J^{n+1}$ is independent of $\Theta$.
Therefore, the incremental minimizer sets are independent of $\Theta$.
When the minimizer is unique at every step, the resulting history
$h_{\rm inc}$ is also independent of $\Theta$.

Consequently, the history-level Gibbs-type measure concentrates on $h_{\rm inc}$ only when $h_{\rm inc}$ is the unique $h_{\rm MAP}$ under the conditions established in the preceding subsection.
Otherwise, the $\Theta \to 0$ limit selects $\mathcal{M}_{\rm MAP}$ rather $h_{\rm inc}$.

The constructions of $h_{\rm cont}$, $h_{\rm MAP}$, $h_{\rm inc}$, and $h_{\rm GSM}$, and their conditional relationships are summarized in Figure~\ref{fig2}.

\section{Representative forward-in-time problems and comparison of histories}\label{Xsec16-4}\label{sec:theta0}

Seven representative forward-in-time problems with a prescribed initial state $z^0$ are considered in this section.
For each problem, the constrained path functional $\mathcal J$ with GSM-type incremental energy--dissipation constitutive structure is first constructed, after which the stationarity conditions for the four histories introduced in section~\ref{sec:GSM} are derived and compared: the continuous minimizing history $h_{\rm cont}$, the discrete MAP history $h_{\rm MAP}$, the causal incremental history $h_{\rm inc}$, and the GSM history $h_{\rm GSM}$.
For each of the seven cases, the framework shows that under the conditions when $h_{\rm MAP} = h_{\rm cont}$ and $h_{\rm inc} = h_{\rm GSM}$, in general,
\begin{equation}\label{eqn:maxwellhrel}
h_{\rm MAP} = h_{\rm cont} \neq h_{\rm inc} = h_{\rm GSM},
\end{equation}
and consequently the energy--dissipation cost incurred by them follows,
\begin{equation}\label{eqn:maxwellJrel}
\mathcal{J}(h_{\rm MAP}) = \mathcal{J}(h_{\rm cont}) < \mathcal{J}(h_{\rm inc}) = \mathcal{J}(h_{\rm GSM}).
\end{equation}

This hierarchy, however, should not be interpreted as a deficiency of the causal incremental formulation (including GSM), but as the additional history-level cost incurred by its causal sequential construction.

For clarity, all examples are presented in a small perturbation, spatially one-dimensional scalar setting.
Spatial derivatives such as gradients or Laplacians should be interpreted accordingly.
The first example is developed in more detail than the rest, which follow the same reasoning.

\subsection{Linear viscoelasticity: Maxwell model}\label{Xsec17-4.1}\label{sec:maxwell}

At time $t^n$, the prescribed field $q^n$ is taken as the total strain $\varepsilon^n$, while the internal variable $z^n$ is the viscoelastic strain $\varepsilon^{v,n}$.
The incremental viscoelastic strain rate $ r^{n+1} := \left(\varepsilon^{v,n+1} -\varepsilon^{v,n}\right)/{\Delta t^n}$.
For the Maxwell model with elastic modulus $E>0$ and viscosity $\eta>0$, the incremental GSM-type action is
\begin{equation}\label{eqn:maxwell}
\begin{aligned}
\mathcal A^{n+1} &\left( \varepsilon^{v,n+1}, \varepsilon^{v,n}; \varepsilon^{n+1} \right)
= \int_\Omega \left[ \underbrace{ \frac12 E \left( \varepsilon^{n+1} - \varepsilon^{v,n+1}\right)^2}_{\psi^{n+1}} + \underbrace{ \frac{\eta}{2\Delta t^n}\left( \varepsilon^{v,n+1}-\varepsilon^{v,n}\right)^2}_{\Delta t^n\phi^{n+1}}\right]\,\mathrm dx .
\end{aligned}
\end{equation}

The initial viscoelastic strain $\varepsilon^{v,0}$ and the total strain history $(\varepsilon^1,\ldots,\varepsilon^N)$ are prescribed.
No additional observational contributions or restrictions are imposed, so that $\mathcal C=0$.

Therefore, the constrained path functional is
\begin{equation}\label{eqn:Jmaxwell}
\mathcal J(h) = \sum_{n=0}^{N-1} \mathcal A^{n+1} \left(\varepsilon^{v,n+1},\varepsilon^{v,n};\varepsilon^{n+1}\right),
\end{equation}
for
\begin{equation}
h = \left( \varepsilon^{v,0}, \ldots, \varepsilon^{v,N} \right) \in \mathcal H_{\rm ad}(\varepsilon^{v,0}).
\label{Xeqn46-50}
\end{equation}

\subsubsection*{Continuous minimizing and discrete MAP histories}

The functional \eqref{eqn:Jmaxwell} is strictly convex with respect to the unknown history $(\varepsilon^{v,1},\ldots,\varepsilon^{v,N})$.
Consequently, it admits at most one continuous minimizing history $ h_{\rm cont}=\left(\varepsilon_{\rm cont}^{v,0}, \dots, \varepsilon_{\rm cont}^{v,N} \right)$.
For every interior state $\varepsilon_{\rm cont}^{v,n+1}$, with $n=0,\ldots,N-2$, stationarity gives
\begin{equation}\label{eqn:maxwell-cont-interior}
\eta \frac{ \varepsilon_{\rm cont}^{v,n+1} - \varepsilon_{\rm cont}^{v,n} }{ \Delta t^{n}} - \eta \frac{ \varepsilon_{\rm cont}^{v,n+2} - \varepsilon_{\rm cont}^{v,n+1}}{\Delta t^{n+1}} = E \left( \varepsilon^{n+1} - \varepsilon_{\rm cont}^{v,n+1} \right),
\end{equation}
and for the final state,
\begin{equation}\label{eqn:maxwell-cont-final}
\eta \frac{ \varepsilon_{\rm cont}^{v,N} - \varepsilon_{\rm cont}^{v,N-1}}{\Delta t^{N-1}} = E\left(\varepsilon^N-\varepsilon_{\rm cont}^{v,N}\right).
\end{equation}

Thus, each interior state of $h_{\rm cont}$ is determined through a two-way coupling with both the immediately preceding state and the next state.

Assuming that the unique continuous minimizer belongs to the candidate set $\mathcal H(\varepsilon^{v,0})$, the conditions established in section~\ref{sec:MAP} give
\begin{equation}\label{eqn:maxwell-map-cont}
h_{\rm MAP}=h_{\rm cont}.
\end{equation}
The limit $\Theta\to0$ then causes the discrete Gibbs-type measure to concentrate on this history.

\subsubsection*{Causal incremental and GSM histories}

The causal incremental history is constructed by minimizing one incremental action at a time.
The incremental action is strictly convex in $\varepsilon^{v,n+1}$, and its stationarity condition is
\begin{equation}\label{eqn:maxwell-inc}
\eta \frac{ \varepsilon_{\rm inc}^{v,n+1} - \varepsilon_{\rm inc}^{v,n}}{ \Delta t^n } = E \left( \varepsilon^{n+1} - \varepsilon_{\rm inc}^{v,n+1} \right).
\end{equation}

Equation~\eqref{eqn:maxwell-inc} is the classical incremental
Maxwell relation.
Since no observational contribution or restriction is present and the incremental minimizer is unique at every step,
\begin{equation}\label{eqn:maxwell-inc-gsm}
h_{\rm inc}=h_{\rm GSM}.
\end{equation}

\subsubsection*{Comparison of the four histories}

Even though the same incremental Maxwell functionals are used in both constructions, the simultaneous and sequential minimizations do not generally produce the same history.
Indeed, substituting \eqref{eqn:maxwell-inc} into \eqref{eqn:maxwell-cont-interior} leaves the residual $- \eta \left(\varepsilon_{\rm inc}^{v,n+2} - \varepsilon_{\rm inc}^{v,n+1} \right)/ \Delta t^{n+1}$.
Therefore, the causal incremental history satisfies the continuous history-level stationarity equations only under the strong compatibility conditions
\begin{equation}
\varepsilon_{\rm inc}^{v,n+2} = \varepsilon_{\rm inc}^{v,n+1}, \qquad n=0,\ldots,N-2.
\label{Xeqn52-56}
\end{equation}
These conditions are not satisfied during a general viscoelastic evolution.
Hence from \eqref{eqn:maxwell-map-cont}, the relations \eqref{eqn:maxwellhrel} and \eqref{eqn:maxwellJrel} are recovered.
This implies that enforcing causality incurs an additional history-level cost during non-trivial evolution.

This is a crucial result and is further exemplified to highlight its importance.
Consider a spatially homogeneous two-increment problem with a unit domain measure and $E=\eta=\Delta t^0=\Delta t^1=1$, $\varepsilon^{v,0}=0$ along with $ \varepsilon^1=\varepsilon^2=1$.
Sequential Maxwell updates give $h_{\rm inc} = h_{\rm GSM} = \left(0,0.5,0.75 \right)$, whereas simultaneous minimization gives $h_{\rm cont} = \left(0,0.6,0.8 \right)$.
If this continuous minimizer belongs to the candidate set, then $h_{\rm MAP}=h_{\rm cont}$.
The corresponding path functional values are $\mathcal J(h_{\rm cont}) = 0.3$ and $\mathcal J(h_{\rm inc}) = 0.3125$, thus recovering the relations \eqref{eqn:maxwellhrel} and \eqref{eqn:maxwellJrel}.

\subsection{Rate-dependent viscoplasticity: Norton model}\label{Xsec18-4.2}

Similar to section~\ref{sec:maxwell}, the total strain $\varepsilon^n$ is prescribed, but the internal variable is the viscoplastic strain $\varepsilon^{vp,n}$.
With $r^{n+1} = \left(\varepsilon^{vp,n+1}-\varepsilon^{vp,n}\right)/{\Delta t^n}$, the incremental action is
\begin{equation}\label{eqn:norton}
\mathcal A^{n+1} = \int_\Omega \left[ \frac12E \left( \varepsilon^{n+1}-\varepsilon^{vp,n+1} \right)^2 + \Delta t^n\frac{K}{m+1} \left|r^{n+1}\right|^{m+1} \right] \,\mathrm dx ,
\end{equation}
where $K>0$ and $m>0$. The initial state and total-strain history are
prescribed, and $\mathcal C=0$.

The interior states of $h_{\rm cont}$ satisfy
\begin{equation}\label{eqn:norton-cont}
K \left|r_{\rm cont}^{n+1}\right|^{m-1} r_{\rm cont}^{n+1} - K \left|r_{\rm cont}^{n+2}\right|^{m-1} r_{\rm cont}^{n+2} = E \left( \varepsilon^{n+1} - \varepsilon_{\rm cont}^{vp,n+1} \right),
\end{equation}
for $n=0,\ldots,N-2$, while the final-state update is
\begin{equation}
K \left|r_{\rm cont}^{N}\right|^{m-1} r_{\rm cont}^{N} = E \left( \varepsilon^{N} - \varepsilon_{\rm cont}^{vp,N} \right).
\label{Xeqn55-59}
\end{equation}
The path functional is strictly convex; hence, assuming that its unique continuous minimizer belongs to the candidate set, $h_{\rm MAP}=h_{\rm cont}$.

Meanwhile, causal incremental minimization gives the backward-Euler Norton update
\begin{equation}\label{eqn:norton-inc}
K \left|r_{\rm inc}^{n+1}\right|^{m-1} r_{\rm inc}^{n+1} = E \left( \varepsilon^{n+1} - \varepsilon_{\rm inc}^{vp,n+1} \right).
\end{equation}
Since no observational contribution or restriction is present, $h_{\rm inc}=h_{\rm GSM}$.

Comparing \eqref{eqn:norton-cont} with \eqref{eqn:norton-inc} shows that the causal history only satisfies the continuous history-level equations under the strong compatibility conditions
\begin{equation}
K \left|r_{\rm inc}^{n+2}\right|^{m-1} r_{\rm inc}^{n+2} =0 \quad \Leftrightarrow \quad \varepsilon_{\rm inc}^{vp,n+2} = \varepsilon_{\rm inc}^{vp,n+1}, \qquad n=0,\ldots,N-2,
\label{Xeqn57-61}
\end{equation}
These conditions do not hold during general viscoplastic evolution.
Therefore, the relations \eqref{eqn:maxwellhrel} and \eqref{eqn:maxwellJrel} are also obtained here.

\subsection{Rate-independent plasticity}\label{Xsec19-4.3}

The prescribed field remains the total strain $\varepsilon^n$, while the internal variable is now the plastic strain $\varepsilon^{p,n}$.
With $r^{n+1} = (\varepsilon^{p,n+1}-\varepsilon^{p,n})/\Delta t^n$,
\begin{equation}\label{eqn:rate-independent}
\mathcal A^{n+1} = \int_\Omega \left[\frac12 E \left( \varepsilon^{n+1}-\varepsilon^{p,n+1}\right)^2 + \Delta t^n\sigma_y \left|r^{n+1}\right|\right] \,\mathrm dx ,
\end{equation}
where $\sigma_y>0$.
Initial state and total strain history are prescribed, and $\mathcal C=0$.

For the continuous minimizing history, there exist $\xi_{\rm cont}^{n+1} \in \sigma_y\partial|r_{\rm cont}^{n+1}|$ such that the interior states satisfy
\begin{equation}
0 = E\left( \varepsilon_{\rm cont}^{p,n+1} - \varepsilon^{n+1} \right) + \xi_{\rm cont}^{n+1} - \xi_{\rm cont}^{n+2}, \qquad n=0,\ldots,N-2,
\label{Xeqn59-63}
\end{equation}
while the final-state update is
\begin{equation}
0 = E\left( \varepsilon_{\rm cont}^{p,N} - \varepsilon^N \right) + \xi_{\rm cont}^{N}.
\label{Xeqn60-64}
\end{equation}
The path functional is strictly convex; hence, assuming that its unique continuous minimizer belongs to the candidate set, $h_{\rm MAP}=h_{\rm cont}$.

Meanwhile, for the causal incremental minimization, there exists $\xi_{\rm inc}^{n+1} \in \sigma_y\partial|r_{\rm inc}^{n+1}|$ such that
\begin{equation}\label{eqn:rate-independent-inc}
0 = E\left(\varepsilon_{\rm inc}^{p,n+1}- \varepsilon^{n+1} \right) + \xi_{\rm inc}^{n+1},
\end{equation}
which is the classical rate-independent plasticity update.
Since observations are absent, $h_{\rm inc}=h_{\rm GSM}$.

A sufficient strong compatibility condition for the causal history to satisfy the
continuous history-level equations is
\begin{equation}
\xi_{\rm inc}^{n+2} = 0 \quad\Longleftrightarrow\quad r_{\rm inc}^{n+2}=0 \quad\Longleftrightarrow\quad \varepsilon_{\rm inc}^{p,n+2} = \varepsilon_{\rm inc}^{p,n+1},
\label{Xeqn62-66}
\end{equation}
for $n=0,\ldots,N-2$.
These conditions do not hold during general plastic evolution.
Therefore, the relations \eqref{eqn:maxwellhrel} and \eqref{eqn:maxwellJrel} are also obtained here.

\subsection{Gradient-regularized damage}\label{Xsec20-4.4}

The prescribed field is the total strain $\varepsilon^n$, while the internal variable is the scalar damage field $d^n\in[0,1]$. With $r^{n+1}=(d^{n+1}-d^n)/\Delta t^n$,
\begin{equation}\label{eqn:damage}
\mathcal A^{n+1} = \int_\Omega \left[ g(d^{n+1})\psi_0(\varepsilon^{n+1}) + \frac{\kappa}{2}|\nabla d^{n+1}|^2 + \frac{\eta}{2\Delta t^n} (d^{n+1}-d^n)^2 \right] \,\mathrm dx ,
\end{equation}
where $g'(d)\leq0$, $\kappa>0$, and $\eta>0$. The initial state and total strain history are prescribed, $\mathcal C=0$, and the constitutive admissible set imposes
$d^n\leq d^{n+1} \leq 1$.

Let $\lambda^{n+1}\geq0$ and $\mu^{n+1}\geq0$ denote the Lagrange multipliers associated with damage irreversibility and the upper bound, respectively, satisfying
\begin{equation}
\lambda^{n+1}(d^{n+1}-d^n)=0,
\qquad
\mu^{n+1}(1-d^{n+1})=0.
\label{Xeqn64-68}
\end{equation}
The interior states of $h_{\rm cont}$ satisfy
\begin{equation}\label{eqn:damage-cont}
\begin{aligned}
0={}&
g'(d_{\rm cont}^{n+1})\psi_0(\varepsilon^{n+1})
-\kappa\Delta d_{\rm cont}^{n+1}
+\eta r_{\rm cont}^{n+1}
-\eta r_{\rm cont}^{n+2}\\
&-\lambda_{\rm cont}^{n+1}
+\lambda_{\rm cont}^{n+2}
+\mu_{\rm cont}^{n+1},
\qquad n=0,\ldots,N-2,
\end{aligned}
\end{equation}
while the final-state update is
\begin{equation}
0=
g'(d_{\rm cont}^{N})\psi_0(\varepsilon^{N})
-\kappa\Delta d_{\rm cont}^{N}
+\eta r_{\rm cont}^{N}
-\lambda_{\rm cont}^{N}
+\mu_{\rm cont}^{N}.
\label{Xeqn66-70}
\end{equation}
These equations are supplemented by the corresponding natural or prescribed boundary conditions. If $g$ is convex, the path functional is strictly convex; hence, assuming that its unique continuous minimizer belongs to the candidate set, $h_{\rm MAP}=h_{\rm cont}$.

Meanwhile, causal incremental minimization gives
\begin{equation}\label{eqn:damage-inc}
0= g'(d_{\rm inc}^{n+1})\psi_0(\varepsilon^{n+1}) -\kappa\Delta d_{\rm inc}^{n+1} +\eta r_{\rm inc}^{n+1} -\lambda_{\rm inc}^{n+1} +\mu_{\rm inc}^{n+1}.
\end{equation}
Since no observations are present, $h_{\rm inc}=h_{\rm GSM}$.

Comparing \eqref{eqn:damage-cont} and \eqref{eqn:damage-inc} shows that the causal history satisfies the continuous history-level equations only under the strong compatibility conditions
\begin{equation}
\eta r_{\rm inc}^{n+2}
=
\lambda_{\rm inc}^{n+2}
\quad\Longleftrightarrow\quad
r_{\rm inc}^{n+2}=0
\ \text{and}\
\lambda_{\rm inc}^{n+2}=0,
\qquad n=0,\ldots,N-2.
\label{Xeqn68-72}
\end{equation}
These conditions do not hold during general damage evolution. Therefore, the relations \eqref{eqn:maxwellhrel} and \eqref{eqn:maxwellJrel} are also obtained here.

\subsection{Non-conserved phase-field solidification: Allen--Cahn type}\label{Xsec21-4.5}\label{sec:allencahn}

The prescribed field is the undercooling $\Delta T^n$, while the internal variable is the scalar order parameter $\eta^n$, distinguishing solid and liquid. With $r^{n+1}=(\eta^{n+1}-\eta^n)/\Delta t^n$, the incremental action is
\begin{equation}\label{eqn:allencahn}
\begin{aligned}
&\mathcal A^{n+1} =
\int_\Omega \left[ \frac{A}{4}\left((\eta^{n+1})^2-1\right)^2 -B\eta^{n+1}\Delta T^{n+1} +\frac{\kappa}{2}|\nabla\eta^{n+1}|^2 +\frac{\zeta}{2\Delta t^n} \left(\eta^{n+1}-\eta^n\right)^2 \right] \mathrm dx ,
\end{aligned}
\end{equation}
where $A,B,\kappa,\zeta>0$. Initial state and undercooling history are prescribed, $\mathcal C=0$, and homogeneous Neumann boundary conditions are imposed.

The interior states of $h_{\rm cont}$ satisfy
\begin{equation}\label{eqn:allencahn-cont}
A\left[
(\eta_{\rm cont}^{n+1})^3-\eta_{\rm cont}^{n+1}
\right]
-B\Delta T^{n+1}
-\kappa\Delta\eta_{\rm cont}^{n+1}
+\zeta r_{\rm cont}^{n+1}
-\zeta r_{\rm cont}^{n+2}
=0,
\end{equation}
for $n=0,\ldots,N-2$, while the final state update is
\begin{equation}
A\left[
(\eta_{\rm cont}^{N})^3-\eta_{\rm cont}^{N}
\right]
-B\Delta T^{N}
-\kappa\Delta\eta_{\rm cont}^{N}
+\zeta r_{\rm cont}^{N}
=0.
\label{Xeqn71-75}
\end{equation}
The path functional is generally nonconvex, so the continuous and discrete MAP minimizers need not be unique.
If $\mathcal M_{\rm cont}\cap\mathcal H\neq\varnothing$, then $\mathcal M_{\rm MAP} = \mathcal M_{\rm cont}\cap\mathcal H$.

Meanwhile, causal incremental minimization gives the backward Euler Allen--Cahn update
\begin{equation}\label{eqn:allencahn-inc}
A\left[
(\eta_{\rm inc}^{n+1})^3-\eta_{\rm inc}^{n+1}
\right]
-B\Delta T^{n+1}
-\kappa\Delta\eta_{\rm inc}^{n+1}
+\zeta r_{\rm inc}^{n+1}
=0.
\end{equation}
Since observations are absent, the causal incremental and GSM constructions have the same incremental minimizer sets; with a unique minimizer at every step, $ h_{\rm inc}=h_{\rm GSM}$.

Comparing \eqref{eqn:allencahn-cont} with \eqref{eqn:allencahn-inc} shows that the causal history only satisfies the continuous history-level equations under the strong compatibility conditions
\begin{equation}
\zeta r_{\rm inc}^{n+2}=0
\quad\Longleftrightarrow\quad
\eta_{\rm inc}^{n+2}
=
\eta_{\rm inc}^{n+1},
\qquad
n=0,\ldots,N-2.
\label{Xeqn73-77}
\end{equation}
These conditions do not hold during general phase-field evolution. Therefore, when the relevant minimizers are unique and the continuous minimizer is represented in the candidate set, then the relations \eqref{eqn:maxwellhrel} and \eqref{eqn:maxwellJrel} are also recovered here.

\subsection{Conserved phase-field solidification: Cahn--Hilliard type}\label{Xsec22-4.6}

As in section~\ref{sec:allencahn}, the prescribed field is the
undercooling $\Delta T^n$, while the internal variable is the order
parameter $\eta^n$. A phase-field flux $j^{n+1}$ is introduced, and the
incremental action is
\begin{equation}\label{eqn:cahnhilliard}
\begin{aligned}
&\mathcal A^{n+1} = \\
&\int_\Omega \left[ \frac{A}{4}\left((\eta^{n+1})^2-1\right)^2 -B\eta^{n+1}\Delta T^{n+1} +\frac{\kappa}{2}|\nabla\eta^{n+1}|^2 +\frac{\Delta t^n}{2M}|j^{n+1}|^2 \right] \mathrm dx,
\end{aligned}
\end{equation}
where $A,B,\kappa,M>0$, subject to
\begin{equation}\label{eqn:CHconstraint}
\frac{\eta^{n+1}-\eta^n}{\Delta t^n}
+\nabla\cdot j^{n+1}=0.
\end{equation}
The initial state and undercooling history are prescribed, and
$\mathcal C=0$.

Let
\begin{equation}
\mu_{( {\ } )}^{n+1} := A\left[ (\eta_{( {\ } )}^{n+1})^3-\eta_{( {\ } )}^{n+1} \right]-B\Delta T^{n+1} -\kappa\Delta\eta_{( {\ } )}^{n+1}.
\label{Xeqn76-80}
\end{equation}
For the continuous minimizing history, there exist constraint multipliers $\lambda_{\rm cont}^{n+1}$ such that
\begin{equation}\label{eqn:CH-cont}
\mu_{\rm cont}^{n+1} -\lambda_{\rm cont}^{n+1} +\lambda_{\rm cont}^{n+2} =0, \qquad n=0,\ldots,N-2,
\end{equation}
while the final state relation is
\begin{equation}
\mu_{\rm cont}^{N}-\lambda_{\rm cont}^{N}=0.
\label{Xeqn78-82}
\end{equation}
At every increment,
\begin{equation}
j_{\rm cont}^{n+1} = -M\nabla\lambda_{\rm cont}^{n+1}, \qquad \frac{\eta_{\rm cont}^{n+1}-\eta_{\rm cont}^{n}}{\Delta t^n} +\nabla\cdot j_{\rm cont}^{n+1}=0.
\label{Xeqn79-83}
\end{equation}
Since the path functional is generally nonconvex, its minimizers need not be unique.
If $\mathcal M_{\rm cont}\cap\mathcal H\neq\varnothing$, then $\mathcal M_{\rm MAP} = \mathcal M_{\rm cont}\cap\mathcal H$.

Meanwhile, causal incremental minimization gives
\begin{equation}\label{eqn:CH-inc}
\lambda_{\rm inc}^{n+1} = \mu_{\rm inc}^{n+1}, \qquad j_{\rm inc}^{n+1} = -M\nabla\mu_{\rm inc}^{n+1},
\end{equation}
and hence the backward-Euler Cahn--Hilliard update
\begin{equation}
\frac{\eta_{\rm inc}^{n+1}-\eta_{\rm inc}^{n}}{\Delta t^n} = \nabla\cdot \left( M\nabla\mu_{\rm inc}^{n+1} \right).
\label{Xeqn81-85}
\end{equation}
Since $\mathcal C=0$ and both constructions impose the same conservation law, unique incremental minimizers give $h_{\rm inc}=h_{\rm GSM}$.

Comparing \eqref{eqn:CH-cont} with \eqref{eqn:CH-inc} shows that a sufficient strong compatibility condition for the causal history to satisfy the
continuous history-level equations is
\begin{equation}
\mu_{\rm inc}^{n+2}=0,
\qquad n=0,\ldots,N-2.
\label{Xeqn82-86}
\end{equation}
These conditions do not hold during general conserved phase-field evolution.
Therefore, when the relevant minimizers are unique and the continuous minimizer belongs to the candidate set, the relations \eqref{eqn:maxwellhrel} and \eqref{eqn:maxwellJrel} are also recovered here.

\subsection{Onsager-type coupled thermoelectric kinetics}\label{Xsec23-4.7}

The internal variable is the vector $z^n=(T^n,V^n)^{\top}$. With $r^{n+1}=(z^{n+1}-z^n)/\Delta t^n$, the incremental action is
\begin{equation}\label{eqn:onsager}
\mathcal A^{n+1} = \int_\Omega \left[ \frac12 (\partial_x z^{n+1})^{\top} K (\partial_x z^{n+1}) + \frac{\Delta t^n}{2} (r^{n+1})^{\top}\Lambda r^{n+1} \right] \,\mathrm dx ,
\end{equation}
where $K$ and $\Lambda$ are symmetric positive-definite $2\times2$
matrices. The initial state and boundary data are prescribed, and
$\mathcal C=0$.

The interior states of $h_{\rm cont}$ satisfy
\begin{equation}\label{eqn:onsager-cont}
-\partial_x \left( K\,\partial_x z_{\rm cont}^{n+1} \right) + \Lambda r_{\rm cont}^{n+1} - \Lambda r_{\rm cont}^{n+2} = 0, \qquad n=0,\ldots,N-2,
\end{equation}
while the final state update is
\begin{equation}
-\partial_x \left( K\,\partial_x z_{\rm cont}^{N} \right) + \Lambda r_{\rm cont}^{N} = 0.
\label{Xeqn85-89}
\end{equation}
The path functional is strictly convex; hence, assuming that its unique continuous minimizer belongs to the candidate set, $h_{\rm MAP}=h_{\rm cont}$.

Meanwhile, causal incremental minimization gives
\begin{equation}\label{eqn:onsager-inc}
-\partial_x \left( K\,\partial_x z_{\rm inc}^{n+1} \right) + \Lambda r_{\rm inc}^{n+1} = 0.
\end{equation}
Since no observational contribution or restriction is present, $h_{\rm inc}=h_{\rm GSM}$.

Comparing \eqref{eqn:onsager-cont} with \eqref{eqn:onsager-inc} shows that the causal history only satisfies the continuous history-level equations under the strong {compatibility} conditions
\begin{equation}
\Lambda r_{\rm inc}^{n+2}=0 \quad\Longleftrightarrow\quad z_{\rm inc}^{n+2}=z_{\rm inc}^{n+1}, \qquad n=0,\ldots,N-2,
\label{Xeqn87-91}
\end{equation}
where the equivalence follows from the positive definiteness of $\Lambda$.
These conditions do not hold during general coupled thermoelectric evolution.
Therefore, the relations \eqref{eqn:maxwellhrel} and \eqref{eqn:maxwellJrel} are also recovered here.

\section{Inverse inference at finite-$\Theta$}\label{Xsec24-5}\label{sec:inference}

The inverse inference capability of the framework is demonstrated at finite $\Theta$ using a spatially 1D, single-mode Swift--Hohenberg phase-field model with a nonconvex free energy \cite{swifthohenberg1977}; the Swift--Hohenberg model is widely used to study pattern formation.
In the inference problem, the final state is prescribed exactly, whereas only partial information is available about the initial state.
The finite-$\Theta$ posterior is used to infer the unobserved initial and intermediate states and to quantify uncertainty over admissible histories.
The empirical MAP history, $h_{\rm eMAP}$, and posterior statistics are compared with the continuous minimizing history $h_{\rm cont}$.

Meanwhile, the causal incremental constructions $h_{\rm inc}$ and $h_{\rm GSM}$ are not designed to solve an inverse problem.
Unless a separately prescribed initial state happens to generate the observed final state, the resulting causal history does not belong to the admissible set $\mathcal H_{\rm ad}$.
Therefore, $h_{\rm inc}$ and $h_{\rm GSM}$ are not included in this example.

\subsection{1D one-mode periodic Swift-Hohenberg system: Constrained path functional and probabilistic formulation}\label{Xsec25-5.1}\label{sec:1DSH}

The Swift-Hohenberg model is considered on the 1D periodic domain $\Omega=(0,L)$, with zero mean order parameter and a single Fourier mode $k=2\pi/L$.
The order parameter at each discrete time $t^{n+1}$ is
\begin{equation}
\eta^{n+1}(x)=a^{n+1}\cos(kx+\theta^{n+1})
\label{Xeqn88-92}
\end{equation}
where $a^{n+1} \geq 0$ is the modal amplitude, and $\theta^{n+1}\in[0,2\pi)$ is the phase shift.
The restriction on the amplitude removes the redundant representation, $a\cos(kx+\theta)=(-a)\cos(kx+\theta+\pi)$.
When $a^{n+1}=0$, the phase is physically undefined.
The zero-mean constraint is automatically satisfied by this expression.

The Swift-Hohenberg free energy is taken as
\begin{equation}
\Psi_{\text{SH}}^{n+1} = \int_0^L
\left[\frac{1}{2}\eta^{n+1}\big(\rho+(1+\partial_{xx})^2\big)\eta^{n+1} +\frac{1}{4}\left(\eta^{n+1}\right)^4
\right]\,dx,
\label{Xeqn89-93}
\end{equation}
with $\rho<0$ in the patterned regime.

The incremental dissipation potential is chosen of Allen-Cahn type with dissipation coefficient equal to unity,
\begin{equation}
\Phi^{n+1} = \int_{0}^L \frac{1}{2}\left(r^{n+1}\right)^2 dx, \qquad r^{n+1}=\frac{\eta^{n+1}-\eta^n}{\Delta t^n}
\label{Xeqn90-94}
\end{equation}

Since $(1+\partial_{xx})^2\cos(kx+\theta)=(1-k^2)^2\cos(kx+\theta)$,
the one-mode free energy reduces to
\begin{equation}
\Psi_{\text{SH}}^{n+1} = \frac{L}{4}\big[\rho+(1-k^2)^2\big]\left( a^{n+1}\right)^2+\frac{3L}{32}\left(a^{n+1}\right)^4.
\label{Xeqn91-95}
\end{equation}
The absence of phase dependence reflects the translational invariance of the periodic system.
Even though the quadratic term may make the free energy negative over a range of amplitudes, the positive quartic term dominates as $|a|\to\infty$, preventing the energy from decreasing without bound.
Furthermore,
\begin{equation}
\int_0^L (\eta^{n+1}-\eta^n)^2\,dx
=
\frac{L}{2}\Big[
\left(a^{n+1}\right)^2+(a^n)^2-2 a^{n+1} a^n\cos(\theta^{n+1}-\theta^n)
\Big].
\label{Xeqn92-96}
\end{equation}

Thus, the incremental action written in terms of the amplitude and phase shift is
\begin{equation}\label{eqn:SHaction}
\begin{aligned}
\mathcal{A}^{n+1}(a^{n+1},\theta^{n+1},a^n,\theta^n)
=&
\frac{L}{4}\left[\rho+(1-k^2)^2\right]\left(a^{n+1}\right)^2
+\frac{3L}{32}\left(a^{n+1}\right)^4 \\
&+ \frac{L}{4\Delta t^n} \left[ \left(a^{n+1}\right)^2+(a^n)^2-2a^{n+1}a^n\cos(\theta^{n+1}-\theta^n) \right]
\end{aligned}
\end{equation}

Consider now a discrete history with $N+1$ states
\begin{equation}
h=\{(a^0,\theta^0),\dots,(a^N,\theta^N)\},
\label{Xeqn94-98}
\end{equation}
with the final state precisely known $(a^N,\theta^N)=(\hat{a},\hat{\theta})$ such that $\hat{\eta}(x)=\hat{a}\cos(kx+\hat{\theta})$.

The continuous admissible history space is
\begin{equation}
\mathcal H_{\rm ad} = \left\{ h: a^n\geq0,\; \theta^n\in[0,2\pi), \forall n = 0, \dots, N-1 \text{ and } \; (a^N,\theta^N)=(\hat a,\hat\theta) \right\}.
\label{Xeqn95-99}
\end{equation}
For the discrete Gibbs-type construction, the amplitude and phase variables are represented on sufficiently fine numerical grids defining $\mathcal H \subseteq \mathcal H_{\rm ad}$.

The available information indicates that the initial state is liquid-like, which means that its amplitude should be close to 0 but it is not known precisely.
This is encoded through a penalty term
\begin{equation}
\mathcal{C}_{\mathrm{init}}(a^0)=\frac{\alpha}{2}(a^0)^2,
\label{Xeqn96-100}
\end{equation}
with $\alpha>0$. The penalty favors, rather than strictly imposes, a small initial amplitude.
The initial phase $\theta^0$ is unconstrained because the phase becomes physically irrelevant as $a^0\to0$.

The full constrained path action is then
\begin{equation}\label{eq:SHconstrainedA}
\mathcal{J}(h)
=
\sum_{n=0}^{N-1}\mathcal{A}(a^{n+1},\theta^{n+1},a^n,\theta^n)
+\mathcal{C}_{\mathrm{init}}(a^0).
\end{equation}
and the finite-$\Theta$ posterior over admissible histories is defined using \eqref{eqn:gibbs}.

Since the free energy is non-convex in the amplitude $a$, $\mathcal{J}$ may possess multiple local or global minimizers.
Furthermore, histories with different phase evolution may have comparable values of $\mathcal{J}$ because the free energy is independent of the phase, while the phase-dependent contribution to the dissipation is weighted by $a^{n+1}a^n$.
Therefore, phase variations are only weakly penalized when the amplitudes are small.

At finite $\Theta$, every candidate history with finite action has nonzero probability.
The posterior quantifies uncertainty in the unknown initial and intermediate states and assigns probability to competing admissible histories connecting the liquid-like initial condition to the prescribed final state.

This reduced problem retains the essential dissipative, nonconvex energetic and probabilistic features required to demonstrate the inverse inference capability of the proposed framework.

\subsection{Numerical implementations}\label{Xsec26-5.2}

\subsubsection{Computation of the continuous minimizing history}\label{Xsec27-5.2.1}\label{sec:deterministicSH}

The continuous minimizing history is obtained by simultaneous minimization of \eqref{eq:SHconstrainedA} over $\mathcal H_{\rm ad}$.
In the one-mode problem, the phase variables can be eliminated analytically.

From \eqref{eqn:SHaction}, the phase dependent contribution from increment $n+1$ is proportional to
\begin{equation}
-a^{n+1}a^n \cos\left(\theta^{n+1}-\theta^n\right).
\label{Xeqn98-102}
\end{equation}
Since the amplitudes are nonnegative, this contribution is minimized by maximizing the cosine.
Whenever $a^{n+1}a^n>0$, the minimum is attained for
\begin{equation}
\theta^{n+1}-\theta^n = 2p\pi, \qquad p\in\mathbb Z.
\label{Xeqn99-103}
\end{equation}

Since the final phase $\hat{\theta}$ is prescribed, the following minimizing sequence may be chosen
\begin{equation}
\theta^0=\theta^1=\cdots=\theta^{N-1}
=\theta^N=\hat{\theta}.
\label{Xeqn100-104}
\end{equation}

Consequently, the constrained path functional depends only on the amplitude
\begin{equation}\label{eq:SH-amplitude-functional}
\begin{aligned}
\mathcal J_a (a^0,\ldots,a^{N-1}) ={}& \sum_{n=0}^{N-1} \left[ \frac{L}{4} \left( \rho+(1-k^2)^2 \right) (a^{n+1})^2 + \frac{3L}{32} (a^{n+1})^4
 + \frac{L}{4\Delta t^n} (a^{n+1}-a^n)^2 \right] + \frac{\alpha}{2}(a^0)^2,
\end{aligned}
\end{equation}
where $a^N=\hat a$ is fixed.
For the parameter regime considered below, the minimizing amplitudes remain strictly positive, so the amplitude inequality constraints are inactive.
If an amplitude reaches zero, the corresponding stationarity condition must instead be replaced by its Karush--Kuhn--Tucker condition.

Stationarity with respect to the initial amplitude gives
\begin{equation}\label{eq:a0update}
\frac{2\alpha}{L}a^0 + \frac{a^0-a^1}{\Delta t^0} = 0.
\end{equation}

Stationarity with respect to an interior amplitude $a^{n}$ with $n=1,\ldots,N-2$, gives the interior update
\begin{equation}\label{eq:anupdate}
\left[\rho+(1-k^2)^2\right]a^n+\frac34(a^n)^3+\frac{a^n-a^{n-1}}{\Delta t^{n-1}}+\frac{a^n-a^{n+1}}{\Delta t^n}=0.
\end{equation}

Finally, stationarity with respect to $a^{N-1}$ with prescribed $a^N = \hat{a}$ gives the final unknown update
\begin{equation}\label{eq:aN-1update}
\left[\rho+(1-k^2)^2\right]a^{N-1} + \frac{3}{4}\left(a^{N-1}\right)^3 + \frac{a^{N-1}-a^{N-2}}{\Delta t^{N-2}} + \frac{a^{N-1}-\hat a}{\Delta t^{N-1}} = 0
\end{equation}

Equations {\eqref{eq:a0update}}--\eqref{eq:aN-1update} form a coupled system of $N$ nonlinear equations for the $N$ unknown amplitudes $a^0,\ldots,a^{N-1}$, which can be solved simultaneously using Newton's method.

Let $R^0$, $R^n$, and $R^{N-1}$ denote the residuals defined by \eqref{eq:a0update}--\eqref{eq:aN-1update}, respectively. The Jacobian is tridiagonal.
Its nonzero entries for the first row are
\begin{equation}
J^{0,0} = \frac{2\alpha}{L} + \frac{1}{\Delta t^0}, \qquad J^{0,1} = -\frac{1}{\Delta t^0}.
\label{Xeqn105-109}
\end{equation}
For $1 \leq n \leq N-2$
\begin{equation}\label{eq:Jacobian}
\begin{aligned}
J^{n,n-1} &= -\frac{1}{\Delta t^{n-1}},\\
J^{n,n} &= \rho+(1-k^2)^2 + \frac94(a^n)^2 + \frac{1}{\Delta t^{n-1}} + \frac{1}{\Delta t^n},\\
J^{n,n+1} &= -\frac{1}{\Delta t^n},
\end{aligned}
\end{equation}
and for $n=N-1$
\begin{equation}
\begin{aligned}
J^{N-1,N-2} &= -\frac{1}{\Delta t^{N-2}}, \\
J^{N-1,N-1} &= \rho+(1-k^2)^2 + \frac94(a^{N-1})^2 + \frac{1}{\Delta t^{N-2}} + \frac{1}{\Delta t^{N-1}}.
\end{aligned}
\label{Xeqn107-111}
\end{equation}

The coupled system is solved using a damped Newton method with line search \cite{NocedalWright2006}.
Since the amplitude functional is nonconvex, convergence of Newton's method establishes a stationary history but does not by itself guarantee the global minimum.
In the present demonstration, the converged history obtained from the initial guess specified in section~\ref{sec:simsetup} is used as the numerical approximation of $h_{\rm cont}$.
A more exhaustive search over multiple initial guesses could be used to investigate the presence of additional stationary histories, but it is not undertaken here.

\subsubsection{Metropolis-Hastings algorithm for the finite-$\Theta$ probabilistic problem}\label{Xsec28-5.2.2}\label{sec:metropolis}

To characterize posterior statistics and uncertainty beyond the single continuous minimizing history obtained in the previous section, the finite-$\Theta$ probability distribution is sampled using the Metropolis--Hastings algorithm \cite{metropolis1953,hastings1970}.

In contrast to the continuous solution, for which the minimizing phases could be eliminated analytically, the phases are retained as unknown variables in the probabilistic problem.
Therefore, a sampled history is written as $h=\{(a^0,\theta^0),\dots,(a^{N-1},\theta^{N-1}),(\hat a,\hat\theta)\}$ with $(a^N,\theta^N)=(\hat a,\hat\theta)$ held fixed by the final observation.

The Markov chain is initialized from an admissible history
\[
h_{(0)} = \left\{(a^{0}_{(0)},\theta^{0}_{(0)}), (a^{1}_{(0)},\theta^{1}_{(0)}), {\ldots}, (a^{N-1}_{(0)},\theta^{N-1}_{(0)}), (\hat{a},\hat{\theta})\right\}
\]
with $a_{(0)}^0$ chosen close to 0 to reflect the liquid-like state.

At each Monte Carlo step $m$, one unknown time index $n\in\{0,\ldots,N-1\}$ is selected uniformly and a local proposal is generated according to
\begin{equation}
a^{n\prime} = a^n_{(m)}+\xi^a, \qquad \theta^{n\prime} = \left( \theta^n_{(m)}+\xi^\theta \right) \bmod 2\pi .
\label{Xeqn108-112}
\end{equation}
Here, $\xi^a$ and $\xi^\theta$ are drawn from symmetric discrete Gaussian distributions on the numerical amplitude and phase grids, with zero means and standard deviations $\sigma_a$ and $\sigma_\theta$, respectively.
The grid spacings are chosen sufficiently small relative to these standard deviations that the discrete distributions are implemented numerically using standard Gaussian samplers.

Since the constrained path functional is local in time, only the neighboring incremental contributions $\mathcal A(a^{n},\theta^{n},a^{n-1},\theta^{n-1})$ and $\mathcal A(a^{n+1},\theta^{n+1},a^{n},\theta^{n})$ are modified by the new proposition per Monte Carlo step, together with the initial prior term when $n=0$.
This approach of local modification of the history has the advantage over modifying entire histories that the changes in the action are local, acceptance rate remains relatively good, the implementation is simpler and the computation remains tractable.

Denoting differences between the proposed and current
histories by $\delta$, the change in the path functional is
\begin{equation}\label{eq:local-delta-J}
\delta\mathcal J
=
\begin{cases}
\delta\mathcal A^1
+
\delta\mathcal C_{\rm init},
& n=0,
\\
\delta\mathcal A^n
+
\delta\mathcal A^{n+1},
& 1\leq n\leq N-1.
\end{cases}
\end{equation}
Thus, the complete path functional does not need to be recomputed at
every Monte Carlo step.

Denoting by $h_{(m)}$ the current history and by $h'$ the new proposition, the acceptance probability is computed as
\begin{equation*}
\beta_{(m)} = \min\left\{ 1,\frac{P_\Theta (h') q(h_{(m)}|h')}{P_\Theta(h_{(m)})q(h'|h_{(m)})} \right\} = \min\left\{ 1,\exp\left(-\frac{\delta \mathcal{J}}{\Theta}\right) \right\}.
\end{equation*}
with $\delta \mathcal{J} = \mathcal{J}(h') - \mathcal{J}(h_{(m)})$.
The second equality follows because the discrete Gaussian proposals are symmetric on the numerical grids, the phase proposal is wrapped periodically, and the time index is selected uniformly, so that $q(h_{(m)}|h') = q(h'|h_{(m)})$, for admissible proposals.

If $\beta_{(m)} = 1$, then $\delta \mathcal{J} \leq 0$ i.e., $h'$ has lower or equal constrained action than $h$, and it is accepted with probability 1.
If $\beta_{(m)} < 1$, then $\delta \mathcal{J} > 0$ and $h'$ has a higher cost than $h$.
Its acceptance is conditional as follows: (i) a random number $x$ is drawn from a uniform distribution $(0,1)$, (ii) if $x \leq \beta_{(m)}$ then the proposal is accepted and $h_{(m+1)} = h'$, otherwise it is rejected and $h_{(m+1)} = h_{(m)}$.

\begin{algorithm}[h!]
\caption{\color{black}Metropolis--Hastings sampling for the one-dimensional
single-mode Swift--Hohenberg problem}
\label{alg:MHSH}
\begin{algorithmic}[1]
\color{black}
\State \textbf{Input:} $(\hat a,\hat\theta)$, $\Theta$, $\sigma_a$, $\sigma_\theta$, $M$, $b$, $N$, $L$, $\rho$, $\alpha$, $\{\Delta t^n\}_{n=0}^{N-1}$, and the numerical candidate grids
\State \textbf{Initialize:} choose an admissible history $h_{(0)}$
\For{$m=0,\ldots,M-1$}
    \State Set $h'\gets h_{(m)}$
    \State Select $n$ uniformly from $\{0,\ldots,N-1\}$
    \State Draw symmetric Gaussian increments $\xi^a$ and $\xi^\theta$
    \State Propose
    $a^{n\prime}\gets a^n_{(m)}+\xi^a$ and $\theta^{n\prime} \gets \left(\theta^n_{(m)}+\xi^\theta\right)\bmod 2\pi$
    \If{$a^{n\prime}$ is outside the admissible amplitude grid}
        \State Set $h_{(m+1)}\gets h_{(m)}$
    \Else
        \State Replace the $n$th state of $h'$ by
        $(a^{n\prime},\theta^{n\prime})$
        \State Compute $\delta\mathcal J$ using \eqref{eq:local-delta-J}
        \State Set
        $ \beta_{(m)} \gets \min\left\{ 1, \exp\left(-\delta\mathcal J/{\Theta}\right) \right\}$
        \State Draw $x\sim\mathrm{Uniform}(0,1)$
        \If{$x\leq\beta_{(m)}$}
            \State Set $h_{(m+1)}\gets h'$
        \Else
            \State Set $h_{(m+1)}\gets h_{(m)}$
        \EndIf
    \EndIf
\EndFor
\State Discard the first $b$ samples as burn-in
\State \textbf{Output:} retained histories and posterior statistics
\end{algorithmic}
\end{algorithm}

Algorithm~\ref{alg:MHSH} modifies one time state at each Monte Carlo step and evaluates only the corresponding local change in the constrained path functional.
After discarding the burn-in portion, the retained histories are used to estimate the pointwise posterior means and $95\%$ credible intervals of the amplitudes, posterior mean of the phases, and the posterior distribution of $\mathcal J(h)$.

The empirical MAP history is defined as the retained history with the smallest sampled value of the constrained path functional,
\begin{equation}
h_{\rm eMAP} \in \underset{m\geq b}{\arg\min}\; \mathcal J(h_{(m)}).
\label{Xeqn110-114}
\end{equation}
where $b$ is the burn-in length.

$h_{\rm eMAP}$ is an approximation of the $h_{\rm MAP}$ and need not coincide exactly with $h_{\rm MAP}$.
While $\mathcal{M}_{\rm MAP}$ remains independent of $\Theta$ when $\mathcal J$ is independent of $\Theta$, in a finite Monte Carlo computation, $h_{\rm eMAP}$ may nevertheless depend on $\Theta$ because the set of histories visited by the sampler depends on the Gibbs-type probability distribution.
Decreasing $\Theta$ increases the relative probability of candidate histories with lower values of $\mathcal J$ and may therefore cause $h_{\rm eMAP}$ to approach $\mathcal{M}_{\rm MAP}$.
When $\mathcal H$ represents a continuous minimizer with sufficient accuracy, $h_{\rm eMAP}$ may consequently also approach $h_{\rm cont}$.
Such convergence requires that the sampling algorithm adequately explores the relevant low-$\mathcal J$ region of $\mathcal H$, which may be impractical.
Therefore, any dependence of the empirical MAP estimate on $\Theta$ would be a finite-sampling effect rather than a change in $\mathcal{M}_{\rm MAP}$.

\subsection{Simulation setup}\label{sec:simsetup}

The parameters for the domain size, Swift-Hohenberg energy, initial and final conditions, temporal discretization, Newton solver, and Metropolis-Hastings algorithm are shown in Table~\ref{tab:1}.
The domain length $L=2\pi$ gives $k=1$, so that $\rho = -0.5$. This gives a negative quadratic coefficient, which combined with the positive quartic term results in a non-convex energy.

The initial guess for the amplitude history used for both simultaneous optimization using Newton's method and the Markov-chain initialization is a linear interpolation between
\begin{equation*}
a^0_{(0)}=10^{-6}
\qquad\text{and}\qquad
a^N_{(0)}=\hat a=1.
\end{equation*}

The damped Newton method with line search and the Metropolis Hastings algorithm have been numerically implemented using Python 3.

The damped Newton iterations are terminated when $ \left\|R\right\|_\infty \leq 10^{-10}$, where $R$ denotes the combined amplitude residual vector.

\begin{table}
\caption{Simulation parameters}\label{tab:1}
\begin{center}
\begin{tabular}{ |c|c|c|c| } 
 \hline
 1D one-mode Swift- & Value & Metropolis-Hastings & Value \\ 
 Hohenberg problem & &  & \\
 \hline
 $L$ & $2\pi$ & $\sigma_a$ & 0.05 \\
 \hline
 $r$ & $-0.5$ &
 $\sigma_{\theta}$ & $0.05$ \\
 \hline
 $\Theta$ & $0.05$ & M & $5 \times 10^7$ \\
 \hline
 $\alpha$ & $1000.0$ & Running average window & 100 \\
 \hline
 $\hat{a}$ & $1.0$  & Burn-in ($b$) & $5 \times 10^6$ \\
 \hline
 $\hat{\theta}$ & $0.5$ & Autocorrelation cutoff $L_\ell$ & $3 \times 10^6$ \\
 \hline
 $dt$ & $0.1$ & Random seed & 42 \\
 \hline
 $N$ & $20$ & Newton solver residual tolerance & $10^{-10}$ \\
 \hline
\end{tabular}
\end{center}
\end{table}

For the Metropolis--Hastings algorithm, the initial phase history is chosen as
\begin{equation}
\theta^n_{(0)}=\hat\theta, \qquad n=0,\ldots,N.
\label{Xeqn111-115}
\end{equation}
The burn-in iterations are \textit{a priori} unknown.
The are estimated empirically using a preliminary run of $M=5 \times 10^7$ local Monte Carlo proposals with the parameters shown in Table~\ref{tab:1}.
The proposal acceptance rate is $\sim 0.4$, which is considered good in that it avoids near-complete rejection and near-complete acceptance.
Sampling efficiency is assessed separately through autocorrelation and effective sample-size estimates.

\begin{figure}[!ht]
\centering
\includegraphics[width=0.65\textwidth]{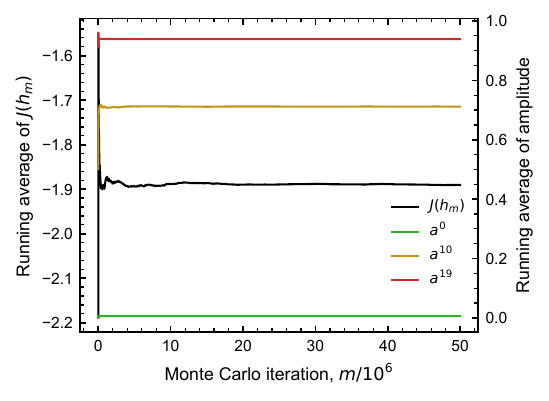}
\caption{\textcolor{black}{Running averages of $\mathcal{J}(h_{(m)})$, $a^0$, $a^{10}$ and $a^{19}$ computed using a window of 100 Monte Carlo proposals.}}\label{fig:figure3}
\end{figure}

The constrained path functional $\mathcal{J}(h_{(m)})$ together with the amplitudes at the initial $a^0$, intermediate $a^{10}$, and penultimate $a^{19}$ time levels were selected as representative global and local observables for testing convergence and sampling efficiency.
Figure~\ref{fig:figure3} reveals that the running average of $\mathcal{J}(h_{(m)})$ exhibits a drift for the first $\sim 5\times 10^6$ proposals prior to stabilizing to an approximately constant value; meanwhile, the selected amplitudes stabilize much earlier.
Therefore, a burn-in length $b = 5\times 10^6$ is {conservatively} chosen for the present Markov chain.

\begin{figure}[!ht]
     \centering
     \begin{subfigure}[b]{0.65\textwidth}
         \centering
         \includegraphics[width=\textwidth]{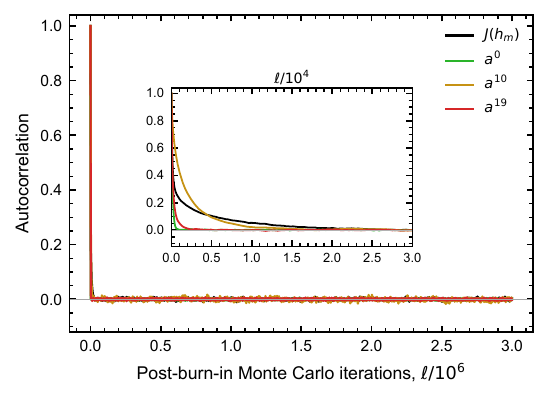}
     \end{subfigure}
     \hfill
     \begin{subfigure}[b]{0.65\textwidth}
         \centering
         \includegraphics[width=0.9\textwidth]{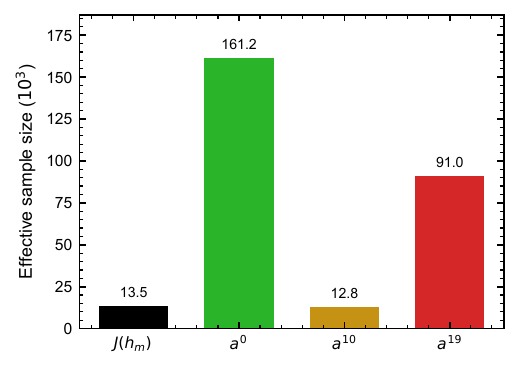}
     \end{subfigure}
     \hfill
        \caption{Autocorrelation as a function of Monte Carlo iterations from retained samples after burnin and corresponding effective sample size (ESS) for $\mathcal{J}(h_{(m)})$, $a^0$, $a^{N/2}$ and $a^{N-1}$.}
        \label{fig:figure4}
\end{figure}

Sampling efficiency after burn-in is assessed using the normalized autocorrelation function and the effective sample size.
Let $\{x_0,x_1,\ldots,x_{M-b-1}\}$ denote the retained sequence of a scalar observable, where $x_i$ represents either $\mathcal J(h_{(m)})$ or one of the selected amplitudes.
Its sample mean is
\begin{equation}
\bar{x}=\frac{1}{M-b}\sum_{i=0}^{M-b-1}x_i.
\label{Xeqn112-116}
\end{equation}
The normalized autocorrelation function is computed as \citep{sokalMonteCarloMethods1997}
\begin{equation}
\hat{\rho}(\ell)
=
\frac{\displaystyle\sum_{i=0}^{M-b-1-\ell}(x_i-\bar{x})(x_{i+\ell}-\bar{x})}
{\displaystyle\sum_{i=0}^{M-b-1}(x_i-\bar{x})^2},
\qquad
\ell=0,1,\ldots,L_\ell,
\label{Xeqn113-117}
\end{equation}
where $L_\ell$ is a user-defined maximum separation between retained samples over which the autocorrelation is evaluated.

The integrated autocorrelation time is estimated using
\begin{equation}
\hat{\tau}_{\mathrm{corr}} = 1 + 2 \sum_{\ell=1}^{L^*} {\hat{\rho}}(\ell),
\label{Xeqn114-118}
\end{equation}
where the summation is truncated at the first $L^*$ for which $\hat{\rho}(L^*+1)\le0$ \citep{geyerPracticalMarkovChain1992}.
The corresponding effective sample size (ESS) is estimated as \citep{sokalMonteCarloMethods1997},
\begin{equation}
\mathrm{ESS} = \frac{M-b}{\hat{\tau}_{\mathrm{corr}}}
\label{Xeqn115-119}
\end{equation}

Figure~\ref{fig:figure4} shows that the autocorrelations of $\mathcal{J}(h_{(m)})$ and $a^{10}$ decay to negligible values after approximately $3 \times 10^4$ local proposals iterations.
The autocorrelations of $a^0$ and $a^{19}$ decay at much lower iterations.
Consequently, the ESS of $a^0$ and $a^{19}$ are much larger than those for $\mathcal{J}(h_{(m)})$ and $a^{10}$ (Figure~\ref{fig:figure3} bottom).
The smallest estimated ESS among the representative observables is $\sim 1.28 \times 10^4$, which is still a substantial ESS for the posterior statistics presented below.

\subsection{Results}\label{Xsec30-5.4}

\subsubsection{Finite-$\Theta$ posterior statistics of retained histories}\label{Xsec31-5.4.1}

Following the analysis performed in section~\ref{sec:simsetup}, the Metropolis-Hastings algorithm is run for $5\times 10^7$ Monte-Carlo proposals using the same parameters shown in Table~\ref{tab:1}.
Posterior statistics are computed on $4.5 \times 10^7$ samples retained after discarding the estimated burn-in.
The simulation and statistics computation takes approximately 17.4 min of wall-clock time on a single core of an Apple M2 pro processor and yields the same proposal acceptance rate of 0.4 as obtained during the test simulation.
Meanwhile, the Newton's method converged in a few seconds after nine damped Newton iterations and gives the numerical approximation of $h_{\rm cont}$.
As discussed previously, the nonconvexity of the path functional means that this single Newton solution is not by itself a proof of global minimality.

\begin{figure}[!ht]
\centering
\includegraphics[width=0.65\textwidth]{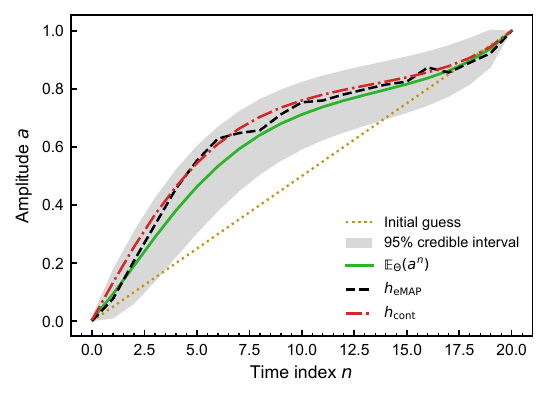}
\caption{\textcolor{black}{Initial guess, 95\% credible interval, posterior mean (expectation $\mathbb{E}_\Theta(a^n)$), and the simultaneously optimized amplitudes for the empirical MAP ($h_{\rm eMAP}$) and continuous ($h_{\rm cont}$) retained admissible histories ($4.5\times 10^7$) with $\Theta = 0.05$.}}\label{fig:figure5}
\end{figure}

Figure~\ref{fig:figure5} compares the initial guess, the 95\% credible interval, the posterior mean (expectation $\mathbb{E}_\Theta (a^n)$), $h_{\rm cont}$ and $h_{\rm eMAP}$ for the amplitude history.
$h_{\rm cont} $evolves smoothly and nonlinearly between the strongly penalized liquid-like initial state and the exactly prescribed final state.
Even though the initial guess is far from the inferred evolution, the posterior statistics are not governed by the initialization of the Markov chain.
For each amplitude $a^n$, the 95\% credible interval is defined by the 2.5\% and 97.5\% posterior quantiles and quantifies the uncertainty associated with the corresponding state.
The interval is narrow near the strongly penalized initial state and becomes zero at the prescribed final state.
It is widest near the middle of the amplitude history, where endpoint observations have the least influence.
$h_{\rm cont}$ lies entirely within the 95\% credible interval and it is therefore compatible with the $\Theta = 0.05$ posterior distribution.
The empirical MAP history $h_{\rm eMAP}$ closely follows the evolution of $h_{\rm cont}$, even though small fluctuations remain because it is the lowest-action history encountered within a finite Monte Carlo sample rather than being the exact discrete MAP minimizer, as explained at the end of section~\ref{sec:metropolis}.
$\mathbb{E}_{\Theta}(a^n)$ is also smooth and nonlinear but visibly differs from $h_{\rm cont}$ and $h_{\rm MAP}$ at intermediate time levels.
This difference is expected because the posterior mean averages over the entirety of sampled histories, whereas $h_{\rm cont}$ and $h_{\rm MAP}$ characterize the low-action region of the posterior.

\begin{figure}[!ht]
\centering
\includegraphics[width=0.65\textwidth]{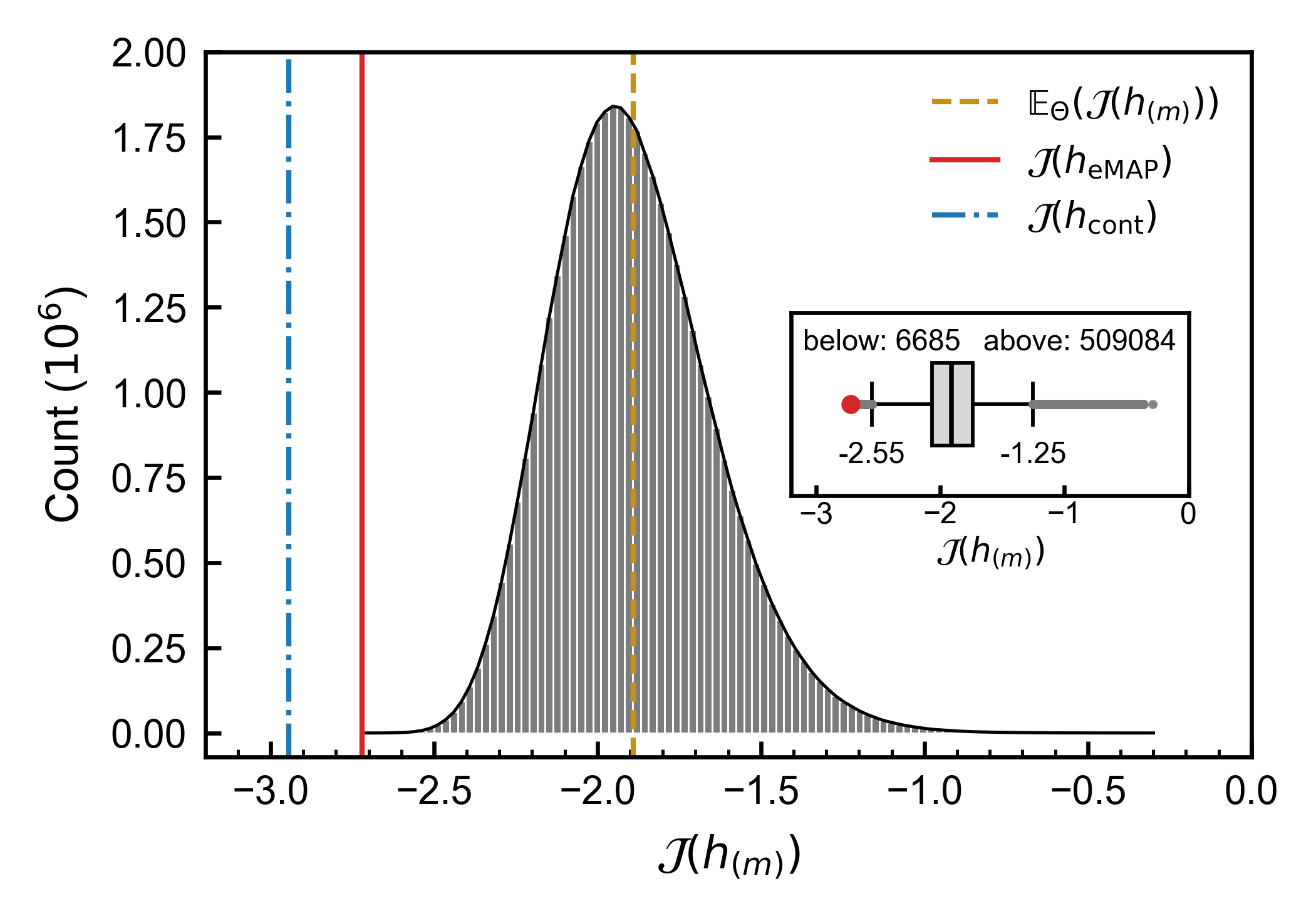}
\caption{\textcolor{black}{Posterior distribution of $\mathcal{J}(h_{(m)})$ obtained from $4.5 \times 10^7$ retained histories at $\Theta = 0.05$. 
The vertical lines indicate $\mathcal J(h_{\rm eMAP})$, the posterior mean action $\mathbb E_\Theta[\mathcal J]$, and $\mathcal J(h_{\rm cont})$.
The inset shows the corresponding box plot, with the whiskers extending to the most extreme non-outlying observations according to Tukey's $1.5$ IQR criterion. 
The numbers of observations below and above the whiskers are also reported.}}\label{fig:figure6}
\end{figure}

Figure~{\ref{fig:figure6}} shows the posterior distribution of $\mathcal{J}(h_{(m)})$.
The distribution is unimodal with a mean of approximately $-1.89$ and median of approximately $-1.91$, and exhibits a right-skewed tail towards higher-$\mathcal{J}$ histories.
By definition, $h_{\rm eMAP}$ attains the smallest $\mathcal{J}$ ($-2.72$) among the retained histories and lies to the low-action end of the posterior distribution.
Thus, it is not representative of the bulk of the posterior mass; which is concentrated at moderately higher $\mathcal{J}$ values.
This gap illustrates the finite-$\Theta$ competition between histories: an individual low-action history has a higher Gibbs-type weight, but the large number of slightly higher action histories collectively carry the bulk of posterior probability.
Therefore, $h_{\rm eMAP}$ is not necessarily a typical posterior history.\vfill\pagebreak

The inset boxplot further demonstrates that the 6685 outlying lowest-action samples are within a mere 6.25\% of the MAP action, indicating the close competition between multiple admissible histories and $h_{\rm eMAP}$.
However, their number should not be interpreted as the number of distinct competing low-$\mathcal{J}$ histories because the retained Markov-chain samples are autocorrelated.

The continuous stationary history $h_{\rm cont}$ attains a lower action than $h_{\rm eMAP}$.
This result is perfectly consistent with $h_{\rm eMAP}$ being only a finite-sample approximation of the discrete MAP history (see last paragraph of section~{\ref{sec:metropolis}}).
Nevertheless, since only one initial guess is employed, the lower value of $\mathcal J(h_{\rm cont})$ does not by itself establish that the computed stationary history is the global continuous minimizer.

\begin{figure}[!ht]
\centering
\includegraphics[width=0.65\textwidth]{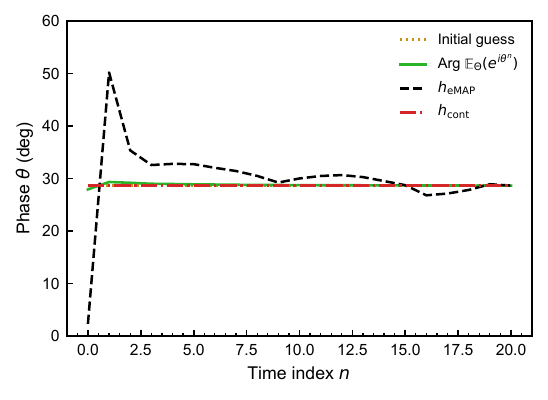}
\caption{\textcolor{black}{Empirical phase evolution as a function of time for retained admissible histories (45 million) with $\Theta = 0.05$. The MAP, posterior mean and deterministic phase evolutions are highlighted along with the initial guess.}}\label{fig:figure7}
\end{figure}

Figure~{\ref{fig:figure7}} compares the initial guess, $h_{\rm cont}$, $h_{\rm eMAP}$, and the posterior mean phase.
The phase enters the constrained action only through the incremental dissipation, which penalizes changes in phase between successive time steps rather than the phase itself.
Therefore, phase variations are weakly penalized near the beginning of the history, where the amplitudes are close to zero, and increasingly penalized as the amplitude grows.

The posterior consequently supports a broad family of phase histories in the initial stages with comparable $\mathcal{J}$.
The 95\% credible interval is not shown because the phase is practically unconstrained when the amplitude is close to zero and it is a periodic variable.
The posterior mean remains nearly constant and agrees closely with the constant phase of $h_{\rm cont}$.
The phase associated with $h_{\rm eMAP}$ exhibits larger early fluctuations but gradually approaches the prescribed final phase as the amplitude increases and phase increments become more costly.

These results demonstrate that the probabilistic framework infers multiple histories compatible with the sparse endpoint information while quantifying uncertainty in the unobserved initial and intermediate states.

\subsubsection{Posterior concentration toward the MAP set as $\Theta\to0$}\label{Xsec32-5.4.2}

\begin{figure}
\centering
\includegraphics[width=\textwidth]{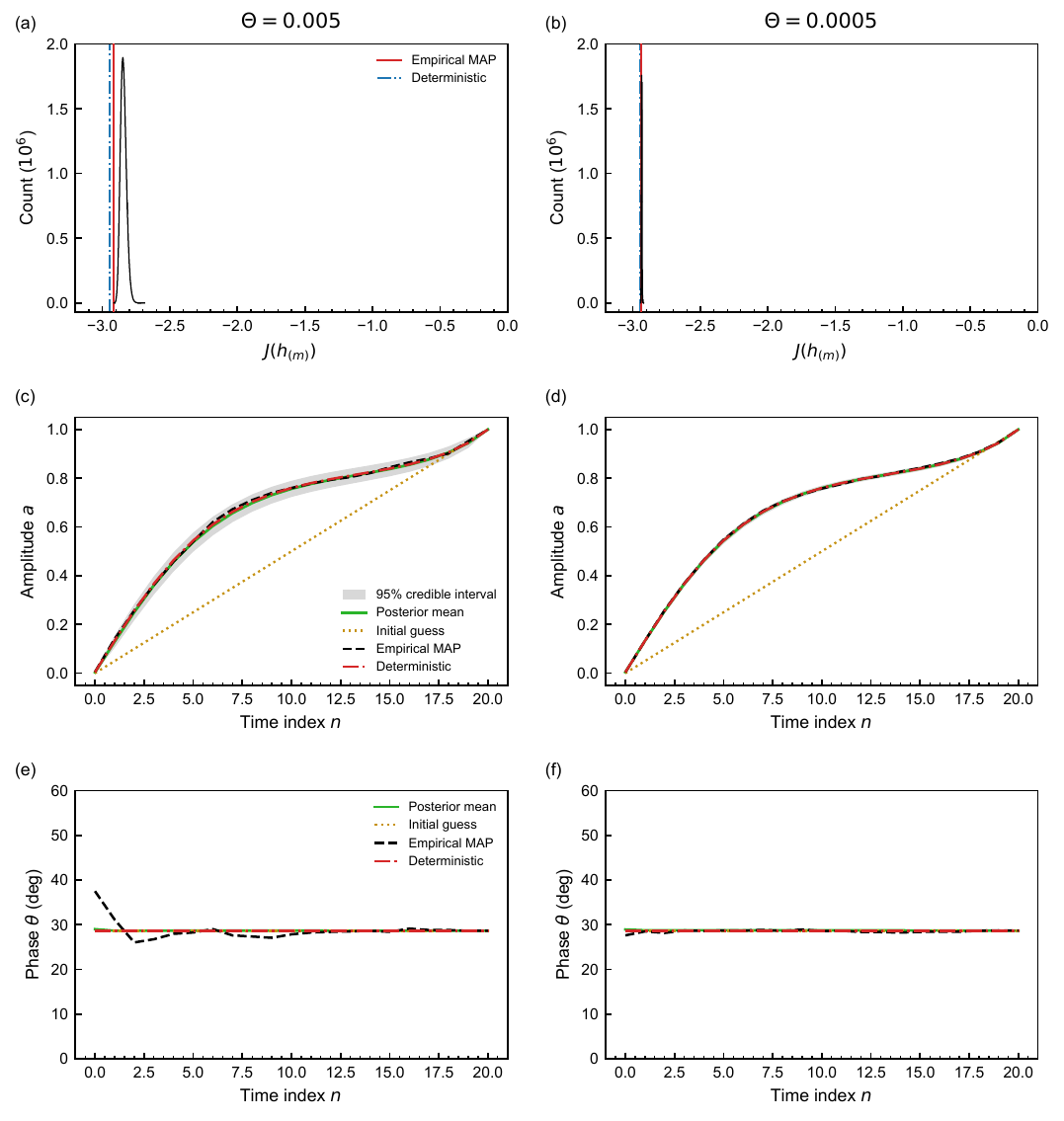}
\caption{Posterior $\mathcal{J}$ distributions and inferred histories for decreasing values of $\Theta$. Panels (a), (c), and (e) correspond to $\Theta=0.005$, whereas panels (b), (d), and (f) correspond to $\Theta=0.0005$. (a) and (b) show the posterior distributions of $\mathcal J$ together with $\mathcal J(h_{\rm eMAP})$ and $\mathcal J(h_{\rm cont})$; (c) and (d) show the amplitude statistics; and (e) and (f) show the phase statistics. Together with Figures~{\ref{fig:figure5}}--\ref{fig:figure7}, these results illustrate the concentration of the posterior toward the low-$\mathcal{J}$ history set as $\Theta$ decreases.}
\label{fig:Figure8}
\end{figure}

To numerically illustrate the posterior concentration as $\Theta\to0$, the inverse inference problem is repeated for $\Theta=0.005$ and $\Theta=0.0005$.
Figure~{\ref{fig:Figure8}} shows the corresponding $\mathcal{J}$ distributions and inferred amplitude and phase histories.

As $\Theta$ decreases, the posterior distribution of $\mathcal{J}$ concentrates towards its low constrained path action edge (Figure~{\ref{fig:Figure8}}(a) and (b)).
At the same time, the 95\% credible interval of the amplitude narrows, while the posterior mean and $h_{\rm eMAP}$ approach the continuous stationary reference $h_{\rm cont}$ (Figure~{\ref{fig:Figure8}}(c) and (d)). 
The phase fluctuations similarly diminish, and the posterior mean and $h_{\rm eMAP}$ approach the constant phase of $h_{\rm cont}$ (Figure~{\ref{fig:Figure8}}(e) and (f)).

For $\Theta=0.0005$, the posterior mean and $h_{\rm eMAP}$ are practically indistinguishable from $h_{\rm cont}$ at the resolution of the plot.
These results provide a numerical illustration of the theoretical concentration of the discrete Gibbs-type measure towards the MAP set as $\Theta\to0$.
Furthermore, the observed approach towards $h_{\rm cont}$ indicates that the numerical candidate set provides a sufficiently fine representation of $h_{\rm cont}$ for the present example.

\section{Conclusion and Perspectives}\label{Xsec33-6}

A probabilistic framework for irreversible kinetics has been introduced in which a constrained path functional $\mathcal J$ encodes constitutive physics and available observations, while a discrete Gibbs-type measure, proportional to $\exp{\left(-\mathcal{J}(h)/\Theta\right)}$, assigns probabilities to an at most countable set $\mathcal{H}$ of candidate histories.
The parameter $\Theta$ controls epistemic uncertainty over these histories.
The resulting Gibbs-type measure can be interpreted as a Bayesian posterior over admissible histories.
Forward-in-time evolution and inverse inference are both treated within this framework and they differ only through the prescribed observations, constraints, and admissible history sets.

A difference was established between the continuous minimizing history {$h_{\rm cont}$}, obtained by simultaneous minimization of $\mathcal{J}$ over entire histories in the admissible space {$\mathcal H_{\rm ad}$, and the discrete maximum-a-posteriori (MAP) history $h_{\rm MAP}$, obtained by simultaneous minimization over $\mathcal H\subseteq\mathcal H_{\rm ad}$.
When a continuous minimizer is represented in $\mathcal H$, the two minimizing sets coincide on that intersection.
Assuming that $\mathcal J$ is independent of $\Theta$, the exact discrete MAP set is also independent of $\Theta$.}
As $\Theta\to0$, the Gibbs-type measure concentrates on this set.

The specialization to generalized standard material (GSM)-type incremental energy--dissipation functionals has revealed an important difference between simultaneous history optimization, and causally incremental evolution.
Even though the simultaneous minimization and causal incremental constructions use the same GSM-type incremental energy--dissipation functionals, the history $h_{\rm inc}$ generated from causal incremental evolutions with a known previous state, and its GSM specialization in the absence of observations $h_{\rm GSM}$, are generally only incrementally optimal.
When the relevant minimizers are unique and $h_{\rm cont}$ is represented in $\mathcal{H}$, the following relations generally hold as shown through seven distinct and representative forward-in-time examples:
\begin{align*}
&h_{\rm MAP} = h_{\rm cont} \neq h_{\rm inc} = h_{\rm GSM}, \\
&\mathcal J(h_{\rm MAP}) = \mathcal J(h_{\rm cont}) < \mathcal{J}(h_{\rm inc}) = \mathcal J(h_{\rm GSM}).
\end{align*}

Thus, if the causal GSM evolution is regarded as the physically representative constitutive history, then the corresponding evolution does not generally follow the lowest energy--dissipation cost over the entire history.
However, this result does not contradict GSM's thermodynamic consistency, but reflects the difference between causal sequential minimization and simultaneous optimization of the entire history.
Furthermore, even though simultaneous minimization is performed over entire histories and connects each state with the previous and next states in time, it does not result in time-symmetric Hamiltonian dynamics; the incremental contributions defining $\mathcal J$ make the histories $h_{\rm cont}$ and $h_{\rm MAP}$ temporally directed.

This difference by itself makes the proposed probabilistic framework more than a description of uncertainty around a deterministic constitutive trajectory.
It provides a history-level inference method that ranks entire admissible paths and need not select the causal GSM history as its MAP history.

At finite $\Theta$, the framework provides posterior statistics over complete histories.
This capability was numerically demonstrated using a Metropolis-Hastings algorithm applied to a Swift--Hohenberg non-convex energy-based inverse problem with the final state prescribed and only partial information about the initial state.
The posterior was used to quantify uncertainty in the unobserved intermediate states and distinguish the empirical MAP history from the posterior mean and the bulk of the posterior mass.
As $\Theta$ decreased, the posterior concentrated towards its low-$\mathcal J$ region, while the empirical MAP and posterior mean approached the continuous stationary history, thus numerically demonstrating the posterior concentration toward the discrete MAP set.

The proposed framework admits several extensions to be explored.
Future developments should address efficient sampling techniques in high-dimensional history spaces, and continuous Gibbs-type measures defined with respect to appropriate reference measures.
Incorporating reversible dynamics together with irreversible kinetics by combining incremental dissipation with global-in-time actions could lead to a probabilistic formulation of continuum mechanics.
More broadly, the constrained path functional could itself become an object of inference to deduce material parameters or the shape of free energy and dissipation potentials.

\appendix

\color{black}
\section{Analogy with GPS routing}\label{app:analogy}

Imagine a scenario where several of us (objective observers) are planning to drive from the same departure location to the same destination. 
Let us further imagine that we are all equipped with identical GPS navigation systems that contain exactly the same road network, traffic information, speed limits and routing algorithm. Before starting our journey, we all enter the same departure and destination locations into our GPS systems. 
Since we are all objective observers with access to exactly the same information, each one of us computes the same cost for any given route between these two locations.

A deterministic navigation system would simply identify one route as the optimal route according to its routing criterion and present only this route to us. 
If we were interested only in reaching our destination, we would be perfectly satisfied with this recommendation and might naturally regard it as the route that should be followed.

Now, let us instead imagine that our GPS system is capable of evaluating not only the optimal route but every admissible route connecting the departure location and destination. 
Knowing the complete road network and assigning each admissible route its corresponding cost, the GPS could rank all admissible routes according to a probability measure that introduces no additional bias beyond the prescribed route cost. 
The optimal route would still be identified as the most probable route, but the GPS would additionally reveal whether several routes are equally optimal, quantify how much less probable the alternative routes are, and compute statistical quantities such as certitude intervals, mean routes and route variability over the ensemble of admissible routes. 
Furthermore, because the same road network and routing criterion apply irrespective of travel direction, the same framework could also rank admissible return journeys under exactly the same prior assumptions.

This is analogous to the probabilistic framework proposed in this work. Rather than assigning probabilities directly to the physical evolution itself, probabilities are assigned to the ensemble of admissible histories through a Gibbs-type measure constructed from a prescribed constrained path functional. The deterministic solution is recovered as the history (or histories) with minimum path cost, while the probabilistic framework additionally quantifies the relative plausibility of all other admissible histories without introducing assumptions beyond those already encoded in the constitutive model and observational constraints.

\bibliographystyle{elsarticle-num} 
\bibliography{references} 

\end{document}